\documentclass[11pt]{article}
\pdfoutput=1
\usepackage{jcappub,amsmath,amssymb}
\usepackage{bm}
\usepackage[table]{xcolor}
\usepackage{graphics}
\usepackage{tikz}

\usetikzlibrary{arrows,patterns,positioning,decorations.markings,decorations.pathmorphing,calc,decorations.pathreplacing}

\tikzset{every node/.append style={minimum size=0.5cm, draw,circle,font=\sffamily\Large\bfseries,inner sep=0.05cm}}%

\tikzset{every loop/.style={}}

\tikzset{every loop/.style={min distance=10mm,in=-35,out=35,looseness=0}}

\pgfdeclarepatternformonly{soft crosshatch}{\pgfqpoint{-2pt}{-2pt}}{\pgfqpoint{8pt}{8pt}}{\pgfqpoint{6pt}{6pt}}%
{
  \pgfsetstrokeopacity{0.2}
  \pgfsetlinewidth{0.1pt}
  \pgfpathmoveto{\pgfqpoint{6.1pt}{0pt}}
  \pgfpathlineto{\pgfqpoint{0pt}{6.1pt}}
  \pgfpathmoveto{\pgfqpoint{0pt}{0pt}}
  \pgfpathlineto{\pgfqpoint{6.1pt}{6.1pt}}
  \pgfusepath{stroke}
}

\usetikzlibrary{arrows,positioning,decorations.markings,decorations.pathmorphing,calc}
\usepackage{todonotes}

\definecolor{hyperref}{RGB}{026,028,087}

\def\gsim{ \lower .75ex \hbox{$\sim$} \llap{\raise .27ex \hbox{$>$}} }
\def\lsim{ \lower .75ex \hbox{$\sim$} \llap{\raise .27ex \hbox{$<$}} }
\def\be{\begin{equation}}
\def\ee{\end{equation}}
\def\bea{\begin{eqnarray}}
\def\eea{\end{eqnarray}}

\newcommand{\ba}{\begin{array}}
\newcommand{\ea}{\end{array}}
\newcommand{\mn}{\mu\nu}

\newcommand{\commentout}[1]{}

\newcommand{\pa}{\partial}

\newcommand{\cI}{{\cal{I}}}

\newcommand{\comment}[1]{}

\newcommand{\bs}{\begin{split}}

\newcommand{\dof}{{\it dof}}
\newcommand{\eft}{{\it EFT}}
\newcommand{\St}{{St\"uckelberg} }

\def\ba{\begin{eqnarray}}
\def\ea{\end{eqnarray}}

\def\nn{\nonumber}

\def\({\left(}
\def\){\right)}

\definecolor{jn}{RGB}{10, 10, 200} 
\definecolor{js}{RGB}{204, 0, 0} 
\definecolor{pgf}{RGB}{10, 150, 10} 

\newcommand*{\mathcolor}{}
\def\mathcolor#1#{\mathcoloraux{#1}}
\newcommand*{\mathcoloraux}[3]{%
  \protect\leavevmode
  \begingroup
    \color#1{#2}#3%
  \endgroup
}
\newlength{\stheight}
\newcommand\textst[1][fu-grey]{
	\ifmmode\setlength{\stheight}{+1.0ex}
	\else\setlength{\stheight}{+0.5ex}
	\fi
	\bgroup\markoverwith{\textcolor{#1}{\rule[\the\stheight]{2pt}{1.0pt}}}\ULon
} 

\newcommand{\textins}[2][fu-grey]{
	\ifmmode\mathcolor{#1}{#2}
	\else\textcolor{#1}{#2}\@\,
	\fi
}

\newcommand{\w}{\wedge}

\newcommand{\N}{{\cal N}}
\newcommand{\EH}{{\it EH}}

\newcommand{\EI}{{E_{^{\!\hspace{.2mm}(i)}}\!}}

\newcommand{\LI}{{\Lambda_{^{\!\hspace{.2mm}(i)}}\!}}

\newcommand{\fI}{{f_{^{\!\hspace{.2mm}(i)}}\!}}

\def\({\left(}
\def\){\right)}

\graphicspath{{./}}
\notoc

  \tikzstyle{vecArrow} = [thick, decoration={markings,mark=at position
   1 with {\arrow[semithick]{open triangle 60}}},
   double distance=1.4pt, shorten >= 5.5pt,
   preaction = {decorate},
   postaction = {draw,line width=1.4pt, white,shorten >= 4.5pt}]

\begin{document}

\title{On Consistent Kinetic and Derivative Interactions for Gravitons}

\author{Johannes Noller}

\affiliation{Astrophysics, University of Oxford, DWB, Keble Road, Oxford, OX1 3RH, UK} 

\emailAdd{noller@physics.ox.ac.uk}

\abstract{The only known fully ghost-free and consistent Lorentz-invariant kinetic term for a graviton (or indeed for any spin-2 field) is the Einstein-Hilbert term. Here we propose and investigate a new candidate family of kinetic interactions and their extensions to derivative interactions involving several spin-2 fields. These new terms generically break diffeomorphism invariance(s) and as a result can lead to the propagation of 5 degrees of freedom for a single spin-2 field -- analogous to ghost-free Massive Gravity. We discuss under what circumstances these new terms can be used to build healthy effective field theories and in the process establish the `Jordan' and `Einstein' frame pictures for Massive-, Bi- and Multi-Gravity.}


\maketitle
\newpage


\begin{figure*}[htbp] 
\begin{center}
\includegraphics[width=.9\linewidth]{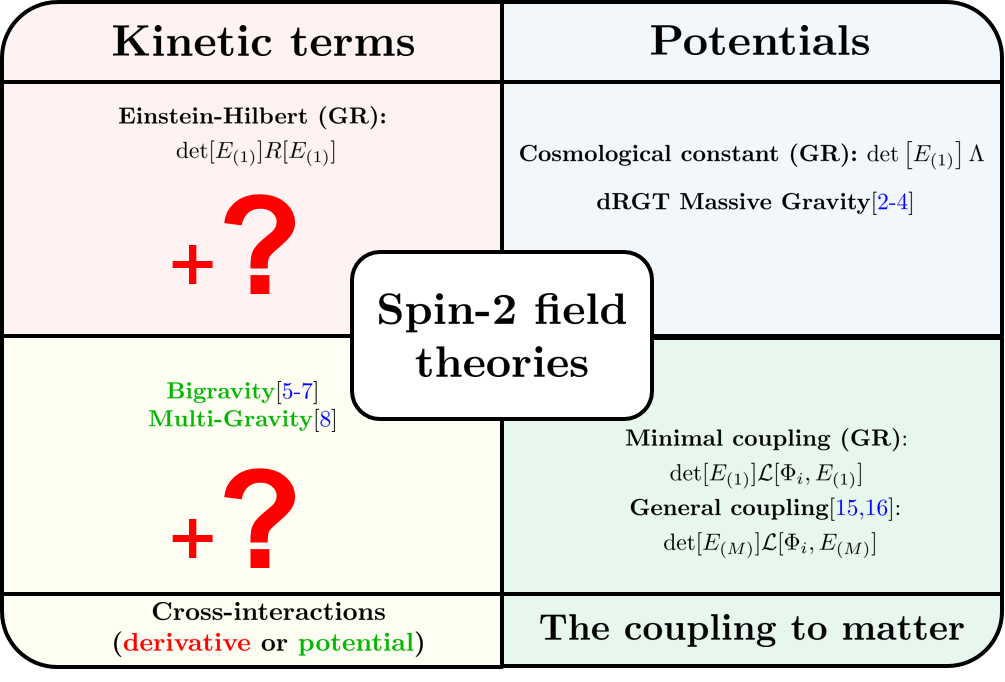}
\end{center}
\caption{Known consistent field theory building blocks for gravitons/spin-2 fields (which are shown in the vielbein picture -- not in the metric picture -- with vielbeins $E_{(i)}$ here). As discussed in section \ref{sec-spin2} such a field theory is in general made up of four types of terms: Kinetic terms for the fields, their potential (non-derivative) interactions, cross-interactions between different fields (if there is more than one dynamical field in the theory, and both of the potential and derivative type) and finally the coupling to the other matter fields $\Phi_i$ in the universe. Note that the `general matter coupling' is only consistent (ghost-free) up to a scale $\Lambda_g > \Lambda_3$. At present the only known fully consistent/ghost-free kinetic term for a graviton/spin-2 field is the Einstein-Hilbert term and there are no known fully consistent/ghost-free derivative interactions involving several spin-2 fields. In this paper we consequently investigate the two big question marks: Whether there are alternative valid kinetic and derivative (cross-)interactions. 
\label{fig-spin2scheme}}
\end{figure*}

\section{Introduction} \label{sec-intro}

The Einstein-Hilbert (henceforth {\it EH}) term is the only known fully ghost-free, consistent and Lorentz invariant derivative interaction (kinetic) term for a single spin-2 field $g$ \cite{deRham:2013tfa}, i.e. also for a graviton. It is fully non-linear, invariant under diffeomorphisms acting on $g$ and propagates the two degrees of freedom (\dof{}) of a massless spin-2 field.
The recent discovery of ghost-free potential-type interactions for a single \cite{deRham:2010ik,deRham:2010kj,Hassan:2011hr}, two \cite{Hassan:2011tf,Hassan:2011zd,Hassan:2011ea} or multiple \cite{Hinterbichler:2012cn} spin-2 fields/gravitons resulted in consistent theories for such fields built out of 1) \EH{} terms for each graviton and 2) the new potential-type interactions.  These non-derivative interactions break some of the diffeomorphism invariance present at the level of the \EH{} terms, leading to the propagation of additional \dof{} (Goldstone bosons of the broken symmetry). E.g. massive gravity generically propagates the five degrees of freedom of a massive spin-2 field.\footnote{It is worth emphasising that broken diffeomorphism invariance(s) can simply be restored in a theory via the \St trick (see \cite{Noller:2013yja} and references therein) -- we are simply using actions with broken/unbroken diffeomorphism invariance as a tool to write down theories with different numbers of propagating \dof, since ultimately what is physical is the number of propagating \dof{} and their interactions, {\it not} the amount of gauge symmetry present in a given action.}

Having constructed consistent theories that break diffeomorphism invariance(s) in their potential (non-derivative) interactions, it is natural to ask whether one can also find kinetic terms other than the \EH{} term for $g$, possibly propagating more than two \dof{}. Various such candidate terms\footnote{Throughout this paper, when we talk about candidate terms, we mean interaction terms, which pass some of the tests to be accepted as consistent interaction terms, but who still remain to be fully understood. To be more specific, this may mean e.g. that the term in question is healthy at the linear level, but a non-linear completion is not yet found (if it exists) or it may mean that the term is healthy non-linearly in some decoupling limit, but it is not understood yet what the range of validity of this term is, i.e. how far beyond the decoupling limit the interactions described by this term can be trusted. Finally it may mean that a certain superposition of interaction terms is healthy, but it is not clear yet, whether arbitrarily detuning the coefficients in this superposition results in a healthy set of interactions as well. We will discuss all of these cases more throughout the paper.} were proposed at the linear level \cite{Hinterbichler:2013eza}(also see \cite{Folkerts:2011ev,Kimura:2013ika,Folkerts:2013mra,Gao:2014jja}), but were shown to have no fully ghost-free non-linear completion \cite{deRham:2013tfa} (there do exist completions which are ghost-free in the decoupling limit, however), so that the \EH{} term still stands as the only known fully ghost-free, non-linear and Lorentz invariant derivative interaction for a spin-2 field. Given that no consistent derivative interactions for several spin-2 fields are known at all (as would be relevant in a Bi- and Multi-Gravity context), the \EH{} term at present is in fact the only fully ghost-free and Lorentz invariant derivative interaction for spin-2 fields full stop. 

In this paper we propose a new candidate family of kinetic and derivative interactions for $\N$ spin-2 fields
\be \label{Der-intro}
{\cal L}_{\rm Der}[E_{(1)},\ldots,E_{(\N)}] = \; {\rm det}\left[\sum_i \alpha_{(i)} E_{(i)}\right]R\left[\sum_i \alpha_{(i)} E_{(i)}\right],
\ee
where the $E_{(i)}$ are vielbeins, $\alpha_{(i)}$ are constant coefficients and $R[X]$ is the Ricci scalar for the effective vielbein $X$. These interactions include new kinetic terms for a single spin-2 field as well as derivative interactions between several fields. We construct these new terms by spotting a set of field re-definitions that leave the overall form of potential self-interactions invariant (only changing coefficients). This allows us to write known fully ghost-free theories of spin-2 fields in a dual formulation that includes the new kinetic and derivative interactions. We also find that the new derivative interactions are closely linked to the general matter coupling proposed by \cite{deRham:2014naa,Noller:2014sta} and as a result derive the Einstein and Jordan frame versions of Massive, Bi- and Multi-Gravity. This enables us to show that the new kinetic and derivative interactions do in fact generically propagate a ghost, which can lie above the decoupling limit, however, meaning that even in the ghostly setups, the new terms can still be part of an interesting valid effective field theory (\eft) below the scale of the ghost. Finally, we extend the theory graph conventions for multi-gravity in light of new matter couplings and derivative interactions in order to aid our understanding of the (symmetry) structure of these theories.
\\

{\it Outline:} This paper is organised as follows. In section \ref{sec-spin2} we  review the vielbein formulation and known consistent field theories of spin-2 fields, collect some useful results along the way and point out a particular property of the known interaction terms under linear field re-definitions of the vielbeins. In section \ref{sec-new} we then use that property to postulate a new set of kinetic and derivative interactions and investigate these individually, with added self-interactions and with added matter coupling terms. In the process we also develop the `kinetic' and `potential' as well as `Einstein' and `Jordan' frame pictures of Massive, Bi- and Multi-Gravity. In section \ref{sec-genspin2} we then consider what generic actions created via superposing the new terms with other known interactions look like. We find that they can (but do not have to) lead to the propagation of a ghost, whose scale can lie above the decoupling limit (again this is true for specific setups). We point out what remains to be done to fully establish the range of validity of generic spin-2 theories containing arbitrary superpositions of the proposed candidate terms (and hence to promote them from ``candidate'' status). Finally we conclude in section \ref{sec-conc}.
\\

{\it Conventions}: Throughout this paper we use the following conventions. $D$ refers to the number of spacetime dimensions and we use Greek letters $\mu, \nu, \ldots$ to denote spacetime indices, which are raised and lowered with the full metric $g_{\mu\nu}$.  Capital Latin letters $A$, $B,\ldots$ are reserved for Lorentz indices and are raised and lowered with the Minkowski metric $\eta_{AB}$. We also use lower case Latin letters $a,b,\ldots$ as a hybrid index, which could be Lorentz or space-time. This is because in perturbative calculations for vielbeins we will contract indices with objects such as $\delta_{\mu}^{\ A}$, so choosing unique labels separating space-time and Lorentz indices is not useful in this case. For vielbein objects the first index is always the space-time index and the second one the Lorentz index, e.g. $E_{\mu}^{\ A}$. Bracketed indices $(i),(j),\ldots$, label the different vielbeins/spin-2 fields -- label indices are not automatically summed over and whether they are upper or lower indices carries no meaning. We denote the completely anti-symmetric epsilon symbol by $\tilde \epsilon$ and define it such that $\tilde\epsilon_{012\cdots D}=1$ regardless of the signature of the metric or the position (up/down) of indices (hence $\tilde\epsilon^{012\cdots D}=\tilde\epsilon_{012\cdots D}=1$). Finally we use commas to denote partial differentiation, e.g. for some tensor $T_\mu$ we have $T_{\mu,\nu} \equiv \partial_\nu T_\mu$.

\section{Consistent spin-2 field theories}\label{sec-spin2}

In this section we quickly summarise known consistent field theories for $\N$ spin-2 fields, collect some useful results along the way and point out some of their properties, which will be relevant for the discussion of new candidate kinetic and derivative interactions.\footnote{For more detailed reviews of the field we refer to  \cite{Hinterbichler:2011tt,deRham:2014zqa}. For a discussion of some of the as yet unresolved issues in the field of massive gravity and whether they are problematic for the field see e.g. \cite{Deser:2014fta} and the counterarguments in \cite{deRham:2014zqa}.} 
We will be working in the vielbein formulation, which in a gravity context was developed in \cite{Nibbelink:2006sz,Hanada:2008hs,Chamseddine:2011mu,Chamseddine:2011bu,Mirbabayi:2011aa}. Note that the metric and vielbein formulations are not always equivalent -- for a related discussion see \cite{Hinterbichler:2012cn,Deffayet:2012zc}. A general action for $\N$ spin-2 fields, parametrised via $\N$ vielbeins $E_{(i)}$ and graphically depicted in figure \ref{fig-spin2scheme}, can be schematically thought of as consisting of the following pieces
\begin{align}\label{schematic}
{\cal S} = \tilde{\cal S}_{\rm der}[E_{(i)},\partial E_{(i)},\ldots] + \tilde{\cal S}_{\rm pot}[E_{(i)}] + {\cal S}_{\rm matter}[\Phi_i, E_{(1)},\ldots, E_{(\N)}].
\end{align}
$\tilde{\cal S}_{\rm der}$ describes the theory's kinetic sector, i.e. derivative interactions (including kinetic terms for the individual fields), $ \tilde{\cal S}_{\rm pot}$ (non-derivative) potential interactions and ${\cal S}_{\rm matter}$ the coupling of the spin-2 fields to the other matter in the universe (the matter fields $\Phi_i$). It is instructive to split up \eqref{schematic} further and write a general action as 
\bea
{\cal S} &=& {\cal S}_{\rm kin}[E_{(i)},\partial E_{(i)},\ldots] + {\cal S}_{\rm der}[E_{(i)},\partial E_{(i)},\ldots] \nn\\
&+& {\cal S}_{\rm pot}[E_{(i)}]   + {\cal S}_{\rm cross}[E_{(i)}] \nn\\
&+& {\cal S}_{\rm matter}[\Phi_i, E_{(1)},\ldots, E_{(\N)}] + {\cal S}_{\rm matter,der}[\Phi_i, E_{(i)},\partial E_{(i)}, \ldots] \label{schematic2}
\eea
The first line now describes derivative interactions for the individual spin-2 fields (kinetic terms ${\cal S}_{\rm kin}[E_{(i)}]$) and derivative interactions between different fields (${\cal S}_{\rm der}$). The second line describes potential (mass) terms for an individual spin-2 field (${\cal S}_{\rm pot}$), e.g. the dRGT ghost-free massive gravity potential, and potential interactions linking several dynamical fields ($ {\cal S}_{\rm cross}$), e.g. ghost-free Bi- \cite{Hassan:2011tf,Hassan:2011zd,Hassan:2011ea} and Multi-Gravity \cite{Hinterbichler:2012cn} potentials.\footnote{The split between terms for the individual fields and cross-terms between different fields is somewhat artificial, in that the `individual' terms can be obtained by considering cross-terms with all but one vielbein being non-dynamical. Nevertheless we will find it useful to consider the two cases separate to begin with.} The final line then describes the general coupling to matter of the spin-2 fields ${\cal S}_{\rm matter}$ \cite{deRham:2014naa,Noller:2014sta} as well as potential derivative couplings to matter ${\cal S}_{\rm matter,der}$. No consistent terms of the form ${\cal S}_{\rm matter,der}$ are known at present.

\subsection{The vielbein formulation}\label{subsec-viel}

We will be working in the vielbein formulation for gravity, where corresponding to each spin-2 field/metric $g$ and its inverse $g^{-1}$ we have a vielbein $E$ and an inverse vielbein $E^{-1}$ satisfying 
\begin{align}
g_{\mu\nu} &= E_{\mu}^{\ A} E_{\nu}^{\ B} \eta_{AB},  &g^{-1}_{\mu\nu} &= E^{-1}{}_{\mu}^{\ A} E^{-1}{}_{\nu}^{\ B} \eta_{AB},
\end{align}
where 
\begin{align}
E^{-1}{}^\mu_{\ A} E_{\nu}^{\ A} &= \delta^\mu_\nu, &E^{-1}{}^\mu_{\ A} E_{\mu}^{\ B} &= \delta^B_A.
\end{align}
Since we will eventually work with several fields/vielbeins, it is imperative to be clear about how indices are raised and lowered at this point. Unless specified otherwise, we will raise/lower space-time/Lorentz indices with the flat Minkowski metric $\eta_{\mu\nu}/\eta_{AB}$. This also means we need to explicitly keep track of when inverse vielbeins are used, since
\begin{align}
E^{-1}{}^\mu_{\ A} g_{\mu\nu}\eta^{AB} & = E_\nu^{\ B}, &E^{-1}{}^\mu_{\ A} \eta_{\mu\nu}\eta^{AB} & \ne E_\nu^{\ B}. 
\end{align}

\subsection{Kinetic terms: The Einstein-Hilbert term} \label{subsec-EH}

The only known fully consistent and ghost-free kinetic term for a single spin-2 field is the Einstein-Hilbert term \cite{deRham:2013tfa}. For a metric $g_{\mu\nu}$ we can build the associated Riemann tensor $R_{abc}^{d}[g]$ and express it in terms of $g$'s affine connection $\Gamma[g]$ as follows
\bea
R_{abc}{}^d[g] = \Gamma [g]^{d}{}_{be} \Gamma [g]^{e}{}_{ac} -  \Gamma [g]^{d}{}_{ae} \Gamma [g]^{e}{}_{bc} -  \Gamma [g]^{d}{}_{bc}{}_{,a} + \Gamma [g]^{d}{}_{ac}{}_{,b},
\eea
where
\bea
\Gamma[g] = \frac{1}{2}g^{-1}{}^{ad}\left(g_{cd,b} + g_{bd,c} - g_{bc,d}   \right).
\eea
Using these expressions and writing $R[g] = R_{\mu\nu\rho}{}^\sigma \delta^\nu_\sigma g^{-1}{}^{\mu\rho}$, we can express the Ricci scalar purely as a function of the vielbein $E$ and its inverse $E^{-1}$
\bea
R[E] &=& \tfrac{3}{2} E^{-1}{}^{ab} E^{-1}{}^{c}{}_{b} E^{-1}{}^{de} E^{-1}{}^{fi} E_{di}{}_{,a} E_{fe}{}_{,c} -  E^{-1}{}^{ab} E^{-1}{}^{c}{}_{b} E^{-1}{}^{de} E^{-1}{}^{fi} E_{de}{}_{,a} E_{fi}{}_{,c}  \nonumber \\ 
&& - 2 E^{-1}{}^{ab} E^{-1}{}^{c}{}_{b} E^{-1}{}^{de} E_{de}{}_{,a}{}_{,c}
-  E^{-1}{}^{ab} E^{-1}{}^{c}{}_{b} E^{-1}{}^{de} E^{-1}{}^{fi} E_{di}{}_{,c} E_{ae}{}_{,f}
\nonumber \\ 
&& + 2 E^{-1}{}^{ab} E^{-1}{}^{c}{}_{b} E^{-1}{}^{de} E^{-1}{}^{fi} E_{fi}{}_{,c} E_{ae}{}_{,d} + \tfrac{1}{2} E^{-1}{}^{ab} E^{-1}{}^{c}{}_{b} E^{-1}{}^{de} E^{-1}{}^{f}{}_{e} E_{fi}{}_{,c} E_{a}{}^{i}{}_{,d} \nonumber \\ 
&& + 2 E^{-1}{}^{ab} E^{-1}{}^{c}{}_{b} E^{-1}{}^{de} E^{-1}{}^{fi} E_{ae}{}_{,c} E_{fi}{}_{,d} + 2 E^{-1}{}^{ab} E^{-1}{}^{c}{}_{b} E^{-1}{}^{de} E_{ae}{}_{,c}{}_{,d}  \nonumber \\ 
&& -  \tfrac{1}{2} E^{-1}{}^{ab} E^{-1}{}^{c}{}_{b} E^{-1}{}^{de} E^{-1}{}^{fi} E_{ci}{}_{,d} E_{ae}{}_{,f} -  E^{-1}{}^{ab} E^{-1}{}^{c}{}_{b} E^{-1}{}^{de} E^{-1}{}^{fi} E_{ae}{}_{,d} E_{ci}{}_{,f} \nonumber \\ 
&& -  \tfrac{1}{2} E^{-1}{}^{ab} E^{-1}{}^{c}{}_{b} E^{-1}{}^{de} E^{-1}{}^{f}{}_{e} E_{a}{}^{i}{}_{,d} E_{ci}{}_{,f} - 2 E^{-1}{}^{ab} E^{-1}{}^{c}{}_{b} E^{-1}{}^{de} E^{-1}{}^{fi} E_{ae}{}_{,c} E_{di}{}_{,f}.\nn\\\label{EHviel}
\eea
We can also express the determinant of $E$ in four dimensions as
\bea
{\rm det}(E) &=& \tfrac{1}{4} E^{ab} E_{b}{}^{c} E_{c}{}^{d} E_{da} -  \tfrac{1}{3} E^{a}{}_{a} E^{bc} E_{c}{}^{d} E_{db} -  \tfrac{1}{8} E^{ab} E_{ba} E^{cd} E_{dc}\nn\\ &+& \tfrac{1}{4} E^{a}{}_{a} E^{b}{}_{b} E^{cd} E_{dc} -  \tfrac{1}{24} E^{a}{}_{a} E^{b}{}_{b} E^{c}{}_{c} E^{d}{}_{d}
\eea
The general expression in D dimensions is ${\rm det}(E)= (1/D!) \epsilon_{A_1,A_2,\ldots,A_D} \epsilon^{\mu_1,\mu_2,\ldots,\mu_D} E_{\mu_1}^{A_1} E_{\mu_2}^{A_2} \ldots E_{\mu_D}^{A_D}$ and it satisfies ${\rm det}(E) = \sqrt{-g}$. 
This then allows us to map the \EH{} term directly from the metric to the vielbein formulation 
\be
{\cal S}_{\rm EH} = M_{Pl}^2 \int d^Dx \sqrt{-g} R[g] \Leftrightarrow M_{Pl}^2 \int d^Dx {\rm det}(E)R[E].
\ee

It will prove to be useful to express the \EH{} term perturbatively, in particular since the graviton is essentially a perturbation $h$ in $E$ around a given background, which we will choose to be flat here for convenience (we also set $M_{Pl}^2 = 1$ in the following). We therefore have
\be
E_{\mu}^{\ A} = \delta_{\mu}^{\ A} + h_{\mu}^{\ A}
\ee
and it is possible to expand the inverse vielbein in terms of $h$ as
\begin{equation}
E^{-1}{}^\mu {}_{A} = \delta^{\mu}{}_{A} - h^{\mu}{}_{A} + h^{\mu d} h_{d A} - h^{\mu d} h_{d c} h^{c}{}_{A} +  h^{\mu d} h_{d}{}^{c} h^{f}{}_{A} h_{c f} + {\cal O}(h^5).
\end{equation}
Note that we may view $\delta_{\mu}{}^{A}$ as the `vielbein' of a non-dynamical Minkowski metric. We will find it useful for comparison later on to compute the full \EH{} term at different orders in $h$. At second order in $h$ one finds (after some integration-by-parts)
\bea
{\cal L}_{\rm EH}^{(2)} &=& - h^{c}{}_{c}{}_{,b} h^{a}{}_{a}{}^{,b} + h^{ab}{}_{,b} h_{a}{}^{c}{}_{,c} + 2 h^{ab}{}_{,a} h_{b}{}^{c}{}_{,c} + 2 h^{a}{}_{a}{}^{,b} h^{c}{}_{b}{}_{,c} -  h_{bc}{}_{,a} h^{ab}{}^{,c} \nonumber \\ 
&& -  \tfrac{5}{2} h_{cb}{}_{,a} h^{ab}{}^{,c} -  \tfrac{3}{2} h_{ac}{}_{,b} h^{ab}{}^{,c} + \tfrac{5}{2} h_{ab}{}_{,c} h^{ab}{}^{,c} -  \tfrac{3}{2} h_{ba}{}_{,c} h^{ab}{}^{,c},\label{EH2}
\eea
and at third order in $h$ the \EH{} term is
\bea
{\cal L}_{\rm EH}^{(3)} &=& -5 h^{ab} h^{cd}{}_{,a} h_{cd}{}_{,b} + 3 h^{ab} h^{cd}{}_{,a} h_{dc}{}_{,b} + 2 h^{ab} h^{c}{}_{c}{}_{,a} h^{d}{}_{d}{}_{,b} + 2 h^{ab} h^{d}{}_{d}{}_{,b} h_{a}{}^{c}{}_{,c} - 2 h^{ab} h^{d}{}_{d}{}_{,b} h^{c}{}_{a}{}_{,c} \nonumber \\ 
&& - 2 h^{ab} h_{b}{}^{c}{}_{,a} h^{d}{}_{d}{}_{,c} - 2 h^{ab} h^{c}{}_{b}{}_{,a} h^{d}{}_{d}{}_{,c} - 2 h^{ab} h^{c}{}_{a}{}_{,b} h^{d}{}_{d}{}_{,c} + 2 h^{ab} h^{d}{}_{d}{}_{,c} h_{ba}{}^{,c}  -  h^{a}{}_{a} h^{d}{}_{d}{}_{,c} h^{b}{}_{b}{}^{,c} \nonumber \\ 
&& - 2 h^{ab} h_{a}{}^{c}{}_{,c} h_{b}{}^{d}{}_{,d} + h^{a}{}_{a} h^{bc}{}_{,c} h_{b}{}^{d}{}_{,d} - 2 h^{ab} h_{b}{}^{c}{}_{,a} h_{c}{}^{d}{}_{,d}  - 2 h^{ab} h^{c}{}_{b}{}_{,a} h_{c}{}^{d}{}_{,d} - 2 h^{ab} h_{a}{}^{c}{}_{,b} h_{c}{}^{d}{}_{,d} \nonumber \\ 
&& + 2 h^{a}{}_{a} h^{bc}{}_{,b} h_{c}{}^{d}{}_{,d} - 2 h^{ab} h_{ab}{}^{,c} h_{c}{}^{d}{}_{,d} - 2 h^{ab} h_{a}{}^{c}{}_{,c} h^{d}{}_{b}{}_{,d} + 2 h^{ab} h_{b}{}^{c}{}_{,a} h^{d}{}_{c}{}_{,d} - 2 h^{ab} h^{c}{}_{b}{}_{,a} h^{d}{}_{c}{}_{,d} \nonumber \\ 
&& + 2 h^{ab} h_{a}{}^{c}{}_{,b} h^{d}{}_{c}{}_{,d}  - 2 h^{ab} h_{ab}{}^{,c} h^{d}{}_{c}{}_{,d} - 2 h^{ab} h_{ba}{}^{,c} h^{d}{}_{c}{}_{,d} + 2 h^{a}{}_{a} h^{b}{}_{b}{}^{,c} h^{d}{}_{c}{}_{,d} + 2 h^{a}{}_{b} h_{d}{}^{c}{}_{,c} h_{a}{}^{b}{}^{,d} \nonumber \\ 
&& -  h^{ab} h_{cd}{}_{,b} h_{a}{}^{c}{}^{,d} + h^{ab} h_{dc}{}_{,b} h_{a}{}^{c}{}^{,d} + h^{ab} h_{bd}{}_{,c} h_{a}{}^{c}{}^{,d} + h^{ab} h_{db}{}_{,c} h_{a}{}^{c}{}^{,d} - 3 h^{ab} h_{bc}{}_{,d} h_{a}{}^{c}{}^{,d} \nonumber \\ 
&& + 3 h^{ab} h_{cb}{}_{,d} h_{a}{}^{c}{}^{,d} + h^{ab} h_{cd}{}_{,a} h_{b}{}^{c}{}^{,d} + 3 h^{ab} h_{dc}{}_{,a} h_{b}{}^{c}{}^{,d} + 2 h^{ab} h_{da}{}_{,c} h_{b}{}^{c}{}^{,d} -  h^{a}{}_{a} h_{cd}{}_{,b} h^{bc}{}^{,d} \nonumber \\ 
&& -  \tfrac{5}{2} h^{a}{}_{a} h_{dc}{}_{,b} h^{bc}{}^{,d} -  \tfrac{3}{2} h^{a}{}_{a} h_{bd}{}_{,c} h^{bc}{}^{,d} + \tfrac{5}{2} h^{a}{}_{a} h_{bc}{}_{,d} h^{bc}{}^{,d} -  \tfrac{3}{2} h^{a}{}_{a} h_{cb}{}_{,d} h^{bc}{}^{,d}  + 2 h^{ab} h_{cd}{}_{,b} h^{c}{}_{a}{}^{,d} \nonumber \\ 
&& + 2 h^{ab} h_{db}{}_{,c} h^{c}{}_{a}{}^{,d} - 2 h^{ab} h_{cb}{}_{,d} h^{c}{}_{a}{}^{,d} + 3 h^{ab} h_{cd}{}_{,a} h^{c}{}_{b}{}^{,d} + h^{ab} h_{dc}{}_{,a} h^{c}{}_{b}{}^{,d} + 2 h^{a}{}_{b} h_{a}{}^{b}{}_{,c} h_{d}{}^{c}{}^{,d}.\nn\\\label{EH3}
\eea
Finally let us point out that, by itself, the \EH{} term is of course a fully ghost-free and consistent  \eft{} all the way up to the Planck scale $M_{Pl}$.  Also the \EH{} term $M_{Pl}^2 \int d^Dx {\rm det}(E_{(i)})R[E_{(i)}]$ is invariant under local diffeomorphism symmetries $\fI$ and Lorentz transformations $\LI$ for the field $\EI$ 
\be \label{diff}
\EI_{\mu}^{\ A}(x)\rightarrow {\partial \fI^\nu \over \partial  x^\mu}\EI_{\nu}^{\ A}\left(\fI(x)\right), \, \quad\quad\quad \EI_\mu^{\ A}\rightarrow \LI^{A}_{\ B} \EI^{\ B}_\mu\, .
\ee

\subsection{Potential terms: Ghost-free Massive, Bi- and Multi-Gravity}\label{subsec-pot}

\begin{figure}[tp]
\centering
\begin{tikzpicture}[-,>=stealth',shorten >=0pt,auto,node distance=2cm,
  thick,main node/.style={circle,fill=white!30,draw,font=\sffamily\large\bfseries},arrow line/.style={thick,-},barrow line/.style={thick,->},no node/.style={plain},rect node/.style={rectangle,fill=blue!10,draw,font=\sffamily\large\bfseries},red node/.style={rectangle,fill=red!10,draw,font=\sffamily\large\bfseries},green node/.style={circle,fill=green!20,draw,font=\sffamily\large\bfseries},yellow node/.style={rectangle,fill=yellow!20,draw,font=\sffamily\large\bfseries}]

  \node[fill=white!30,thin,draw=black](200){};      
   \path (200) edge[loop below,dashed,-] node[draw=none,left=1pt]{} (300);

 \node[fill=white!30,thin,draw=black](100)[right=3cm of 200]{};
  \node[main node,thin,draw=black] (101) [right of=100] {};
 
  \node[main node,fill=black!100,scale=0.5] (2)  [right=3cm of 101]{};
   \node[main node,thin,draw=black] (1) [above=1cm of 2] {};
  \node[main node,thin,draw=black] (3) [below left=1cm of 2] {};
   \node[fill=white!30,thin,draw=black] (4) [below right=1cm of 2] {};

         \node[draw=none,fill=none](83)[right=0.5cm of 100]{};

     \draw[-,dashed] (100) to (101);
\draw[-,dashed] (1) to (2);
\draw[-,dashed] (4) to (2);
\draw[-,dashed] (3) to (2);

   \node[draw=none,fill=none,rectangle](92)[below of=200]{\rm \normalsize (a) Cosmological constant};
   
   \node[draw=none,fill=none,rectangle](93)[below of=83]{\rm \normalsize (b) Bimetric};
   
   \node[draw=none,fill=none,rectangle](94)[below of=2]{\rm \normalsize (c) Trimetric};
\end{tikzpicture}
\caption{Theory graphs for ghost-free potential interaction terms: Nodes denote the individual spin-2 fields and dashed lines potential interactions. a) A cosmological constant for a single spin-2 field takes on the form of a loop, since it is a self-interaction. b) Ghost-free potential interactions between two spin-2 fields. We have not yet specified whether these fields are dynamical or not, so this is the potential type interaction for {\it both} Massive and Bi-Gravity. c) A Ghost-free interaction between three spin-2 fields of the form \eqref{potential}. Note that the graphs in this figure only depict potential self-interactions -- how kinetic terms and matter couplings are added to theory graphs is displayed in figures \ref{fig-kinetic} and \ref{fig-matter}.} \label{fig-potentials}
\end{figure}
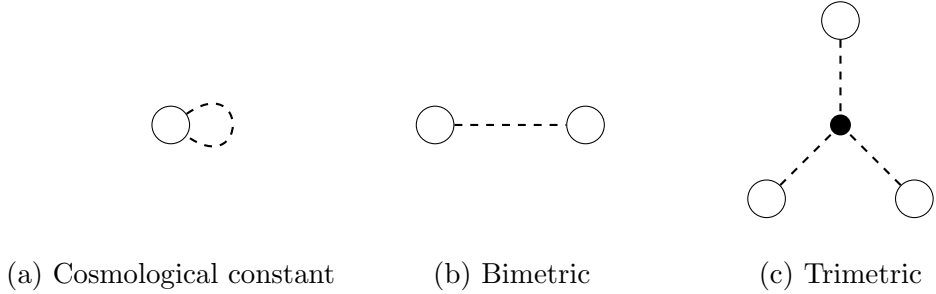

The known ghost-free potential interactions for $\N$ spin-2 fields are those of ghost-free massive gravity \cite{deRham:2010ik,deRham:2010kj,Hassan:2011hr}, Bigravity \cite{Hassan:2011tf,Hassan:2011zd,Hassan:2011ea} and Multi-Gravity \cite{Hinterbichler:2012cn}. In terms of vielbeins they can all be cast in the unified format \cite{Hinterbichler:2012cn} (in $D$ dimensions)
 \bea  
 {\cI}_{(i_1 i_2 \ldots i_D)}
 \equiv  \tilde \epsilon_{A_1 A_2 \cdots A_D} \, \tilde  \epsilon^{\mu_1 \mu_2 \cdots \mu_D} \, 
 {E}_{(i_1)}{}_{\mu_1}^{\ A_1} \,  {E}_{(i_2)}{}_{\mu_2}^{\ A_2} \cdots {E}_{(i_D)}{}_{\mu_D}^{\ A_D},
 \label{density}
\eea
where the indices $(i_1 i_2 \ldots i_D)$ keep track of which fields are interacting.\footnote{If we promote the vielbeins to be proper one-forms ${\bf E}_{(i)}^A=E_{(i)}{}_\mu^{\ A}dx^\mu$, this interaction can be written 
\bea \label{genint}
{\cI}_{(i_1 i_2 \ldots i_D)} d^D x &\equiv & \tilde  \epsilon_{A_1 A_2 \cdots A_D}\, {\bf E}^{A_1}_{(i_1)}\w {\bf E}^{A_2}_{(i_2)}\w \ldots\w {\bf E}^{A_D}_{(i_D)},
\eea
in terms of the usual wedge product. Also note that the anti-symmetric nature of the interaction term means the order of labels in \eqref{density} $(i_1 i_2 \ldots i_D)$ is irrelevant.} Ghost-free massive gravity potential interactions then consist of all the ways to build \eqref{density} with a single dynamical vielbein $E_{(1)}$ and a non-dynamical reference vielbein (in the case of a flat reference metric this non-dynamical vielbein is $E_{(0)}{}_\mu^{\ A} = \delta_\mu^{\ A}$). Ghost-free Bigravity consists of all interactions \eqref{density} that can be build with two dynamical vielbeins $E_{(1)}$ and $E_{(2)}$, and so on. The most general known, fully ghost-free potential interaction for $\N$ spin-2 fields can therefore be written as
\bea \label{potential}
\tilde{\cal S}_{\rm pot}\left[E_{(i)}\right] = \sum_{i_j}^{\N} c_{(i_1 i_2 \ldots i_D)} {\cI}_{(i_1 i_2 \ldots i_D)},
\eea
where the $c_{(i_1 i_2 \ldots i_D)}$ are constant coefficients completely symmetric in all the $i_j$. 
Figure \ref{fig-potentials} depicts the theory graphs for some potentials. We emphasise that all terms of the form \eqref{density} are fully ghost-free individually and that this remains true for any linear superposition of such terms as in \eqref{potential}. One particular such superposition of interaction terms, which we will find to be of relevance later one, is
\bea \label{detform}
\int d^Dx {\rm det}\left[ \sum_i \alpha_{(i)}E_{(i)}{}_\mu^{\ A} \right].
\eea

The potential interaction terms considered here introduce a new scale into the theory. This is the scale of the least suppressed irrelevant operators in the action -- the operators picked out by the so-called decoupling limit. In the case of dRGT ghost-free massive gravity this scale is $\Lambda_3$
\be
\Lambda_{3}=\left(M_{\rm Pl} m^{\lambda-1}\right)^{1/\lambda}= m\left( \frac{M_{\rm Pl}}{m} \right)^{1/ \lambda},
\ee
where $m^2 M_{Pl}^2$ is the coupling constant in front of an interaction term like \eqref{density} (so for \eqref{potential} we schematically have $c_{(i_1,i_2,\ldots,i_D)} = m_{(i_1,i_2,\ldots,i_D)}^2 M_{Pl}^2$). In ghost-free Multi-Gravity this scale is somewhat modified as it picks up a dependence on the number of fields $\N$ -- for details see \cite{Noller:2013yja,deRham:2013awa}. There are further scales of interest which derive from $\Lambda_3$: $\Lambda_Q$, the scale where quantum corrections are no longer suppressed wrt. the tree level result and $\Lambda_V$, the (Vainshtein) scale where classical non-linearities begin to dominate over the linear result. $\Lambda_Q > \Lambda_V$ for ghost-free massive gravity \cite{Hinterbichler:2011tt}, which allows for regions where classical non-linear computations can be trusted (in particular the Vainshtein mechanism) and are not swamped by quantum corrections. More specifically, for ghost-free massive gravity solutions around a source with mass $M$ we can translate the energy scales $\Lambda_Q$ and $\Lambda_V$ into distance (radii) scales $r_Q$ and $r_V$ where quantum corrections/classical non-linearities become important and one finds\cite{Hinterbichler:2011tt} 
\begin{align}
r_Q &\sim \frac{1}{\Lambda_3}  &r_V &\sim \left(\frac{M}{M_{Pl}}\right)^{1/3} \frac{1}{\Lambda_3}.
\end{align}
Also note that $\Lambda_c$, the scale where the \eft{} completely breaks down and e.g. unitarity violations occur, is another in principle distinct (and potentially larger) scale (even though $\Lambda_Q$ to all intents and purposes is the scale where we cannot trust the theory's predictions in the absence of a re-summation for quantum corrections to all orders).

The dRGT ghost-free massive gravity potential (a theory of one dynamical and one non-dynamical vielbein) and in fact any potential of the form \eqref{potential}, where some of the $E_{(i)}$ are non-dynamical, generically break all diffeomorphism symmetries \eqref{diff} that were present in the kinetic sector of the theory\footnote{This is the case if the kinetic sector is simply a linear superposition of \EH{} terms for all dynamical spin-2 fields. We will discuss some exceptions in section \ref{subsec-graph}.}. For a potential like \eqref{potential}, where all the fields are dynamical, the diffeomorphism symmetries for all $E_{(i)}$ \eqref{diff} generically get broken down to their diagonal subgroup (for details see \cite{Hinterbichler:2012cn,Noller:2013yja}).

\subsection{The coupling to matter: Effective matter metrics and vielbeins}\label{subsec-matter}

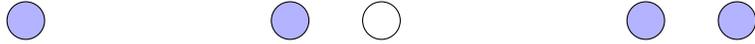
\begin{figure}[tp]
\centering
\begin{tikzpicture}[-,>=stealth',shorten >=0pt,auto,node distance=2cm,
  thick,main node/.style={circle,fill=blue!30,draw,font=\sffamily\large\bfseries},arrow line/.style={thick,-},barrow line/.style={thick,->},no node/.style={plain},rect node/.style={rectangle,fill=blue!10,draw,font=\sffamily\large\bfseries},red node/.style={rectangle,fill=red!10,draw,font=\sffamily\large\bfseries},green node/.style={circle,fill=green!20,draw,font=\sffamily\large\bfseries},yellow node/.style={rectangle,fill=yellow!20,draw,font=\sffamily\large\bfseries}]

  \node[fill=blue!30,thin,draw=black](200){};

 \node[fill=blue!30,thin,draw=black](100)[right=3cm of 200]{};
  \node[fill=white!30,thin,draw=black] (101) [right=0.7cm of 100] {};
 
 \node[fill=blue!30,thin,draw=black](103)[right=3cm of 101]{};
  \node[fill=blue!30,thin,draw=black] (104) [right=0.7cm of 103] {};

         \node[draw=none,fill=none](83)[right=0.25cm of 100]{};
         \node[draw=none,fill=none](85)[right=0.25cm of 103]{};


   \node[draw=none,fill=none,rectangle](92)[below of=200]{\rm \small (a) Minimal coupling};
   
   \node[draw=none,fill=none,rectangle](94)[below of=85]{\rm \small (c) ${\cal L}\left[\Phi_i,E_{(1)} + E_{(2)}\right]$};
   
   \node[draw=none,fill=none,rectangle](93)[below of=83]{\rm \small (b) Bigravity minimal coupling};
   
\end{tikzpicture}
\caption{Theory graphs for different matter couplings: Dark-shaded nodes denote the fields that are directly coupled to matter. a) A single spin-2 field minimally coupled to matter as in \eqref{minimal}. b) Two spin-2 fields, one coupled to matter as in \eqref{minimal} and the other not directly coupled to matter. c) Two spin-2 fields coupled to matter as in \eqref{matterc} or \eqref{matviel}. Note that the graphs in this figure only depict different ways of coupling to matter -- how kinetic terms and potential self-interactions are added to theory graphs is displayed in figures \ref{fig-kinetic} and \ref{fig-potentials}.} \label{fig-matter}
\end{figure}

The standard minimal coupling of a single spin-2 field ($g^{(i)}$ in the metric formalism) to matter 
\be \label{minimal}
\int d^D x \sqrt{-\det \left({\bm g}^{(i)}\right)} {\cal L}\left[\Phi_i, {\bm g}^{(i)} \right]
\ee
is of course fully ghost-free (in the sense that a form of the Boulware-Deser ghost does not get re-exited by this coupling -- naturally one is at liberty to introduce matter ghosts into the matter Lagrangian). Coupling matter minimally to more than one spin-2 field, however, re-exites the Boulware-Deser ghost at a low scale and is consequently not consistent \cite{Yamashita:2014fga,deRham:2014naa}. Recently a more general matter coupling than \eqref{minimal} was proposed in the presence of $\N$ spin-2 fields \cite{deRham:2014naa,Noller:2014sta}.\footnote{In \cite{mattercoupling} we construct the most general matter coupling in the vielbein formulation and investigate whether more general consistent matter couplings can be constructed.} This is of the form
\be \label{matterc}
\int d^D x \sqrt{-\det \left({\bm g}^{(M)}\right)} {\cal L}\left[\Phi_i, {\bm g}^{(M)} \right],
\ee
and consequently obeys the weak equivalence principle by construction. The matter metric ${\bm g}^{(M)}$ can be expressed in terms of the `matter vielbein' $E^{(M)}$ satisfying 
\begin{align} \label{conc-viel}
{\bm g}^{(M)}_{\mn} &= E^{(M)}{}_{\mu}^{\  A} E^{(M)}{}_{\nu}^{\ B} \eta_{AB}, &E^{(M)}{}_{\mu}^{\  A} &= \sum_{i=1}^{\N} \alpha_{(i)} E^{(i)}{}_{\mu}^{\  A}.
\end{align}
As such let us write this new universal coupling to matter in the vielbein language as 
\be \label{matviel}
\int d^D x {\rm det}[E^{(M)}] {\cal L}\left[\Phi_i, E^{(M)} \right].
\ee
Theory graphs for different types of matter couplings are shown in figure \ref{fig-matter}. For some related work exploring models with this new matter coupling further see \cite{Hassan:2014gta,Enander:2014xga,Schmidt-May:2014xla,deRham:2014fha,Gumrukcuoglu:2014xba}.

The new matter coupling \eqref{matterc} generates additional pure spin-2 potential interactions, which are fully ghost-free by construction \cite{deRham:2014naa,Noller:2014sta}. In fact they are of the form \eqref{detform} and come from the cosmological-constant-like piece in \eqref{matviel}, i.e. the constant piece of ${\cal L}$, whether bare or generated by matter loops, gives rise to pure spin-2 interactions. Once non-trivial (i.e. dynamical) matter fields are introduced, however, \eqref{matterc} ceases to be linear in all the lapses of the vielbeins involved \cite{deRham:2014naa}. This means a constraint is lost and the Boulware-Deser ghost re-appears (for a further discussion of this point see \cite{Hassan:2014gta,deRham:2014fha}). The scale of this ghost $\Lambda_g$ lies above $\Lambda_3$, as it can be shown explicitly that the new ghost is not present in the decoupling limit \cite{deRham:2014naa}. So, in a general setting with the new matter coupling, one is dealing with an \eft{} valid up to the scale $\Lambda_g$ or $\Lambda_c$, whichever is lower. At present it is unknown, where between $\Lambda_3$ and $\Lambda_c$ the scale of the matter-coupling ghost $\Lambda_g$ lies.\footnote{If $\Lambda_g$ lies very close to $\Lambda_3$, the theory will essentially have no valid non-linear regime and hence not exhibit the Vainshtein mechanism. If, however, $\Lambda_g$ is parametrically larger than $\Lambda_3$, interesting non-linear phenomena are describable within the range of validity of the \eft.}

\subsection{Invariance under field re-definitions}\label{subsec-inv}

The potential interactions discussed in \ref{subsec-pot} and the general matter coupling discussed in \ref{subsec-matter} above  have an interesting property, which will be key in the following discussion. Namely their form stays invariant under {\it linear} field re-definitions 
\be \label{fieldredef}
E_{(i)} \to \sum_j \beta_{(ij)}E_{(j)}.
\ee
Of course linear field re-definitions performed on a full action leave the physics described invariant. The key observation here is, that field re-definitions of the form \eqref{fieldredef} can be performed on the individual potential and matter coupling terms (rather than on all terms in the full action). This results in a {\it different} theory, i.e. the theory is {\it not} invariant under this change. But the new theory {\it does remain} a valid, consistent, ghost-free \eft{} (up to the scale $\Lambda_g$, should there be a non-trivial matter coupling).

In the following sections we will use this observation in order to generate new candidate kinetic and derivative interactions. Here we point out an interesting analogy, namely that all consistent Massive, Bi- and Multi-Gravity potential interactions could have been discovered in the same fashion. Take a cosmological constant interaction for a single spin-2 field and perform the above field re-definition
\bea \label{ccanalogy}
\int d^Dx {\rm det}\left[E_{(1)}{}_\mu^{\ A} \right] \Lambda \to \int d^Dx {\rm det}\left[ \sum_i \alpha_{(i)}E_{(i)}{}_\mu^{\ A} \right] \Lambda.
\eea
This is the particular superposition of all known potential interaction terms \eqref{detform}. If only one of the fields $E_{(i)}$ is dynamical, these terms will be of the dRGT type, for two dynamical fields we recover the Bigravity interactions and for $\N$ dynamical fields we recover Multi-Gravity interactions. Of course what makes these terms even more interesting is that they are all ghost-free and consistent by themselves, not just in their superposed form \eqref{ccanalogy}. In other words, ghost freedom of \eqref{ccanalogy} is maintained when the coefficients of the individual interactions terms in this superposition are detuned. Nevertheless, finding the right field re-definition could have been used as a neat trick in order to generate all candidate consistent potential interactions in a particular superposition (the coefficients of which could have been subsequently detuned, leading to the finding that {\it any} superposition of these interaction terms is also ghost-free). We will take this approach now with the kinetic sector of the theory, aiming to find new consistent candidate kinetic and derivative interactions.

\section{New kinetic and derivative interactions}\label{sec-new}

In this section we first propose a set of new kinetic and derivative interactions for $\N$ spin-2 fields, using the field re-definitions discussed above. We then discuss their properties, first by themselves, then once potential self-interactions are added and then with an added coupling to matter. This progression is instructive, because superposing two actions, which are ghost-free and consistent by themselves, does not necessarily result in a consistent new theory.\footnote{We thank Claudia de Rham and Andrew Matas for stressing this point in discussions.} We will see an explicit example of this in this section. In addition there are interesting symmetry breaking phenomena when superposing the different interactions, which we will discuss. 

For a general theory with $\N$ spin-2 fields and $\N$ corresponding vielbeins $E_{(i)}$, we propose the following candidate derivative interactions
\be \label{Der}
{\cal L}_{\rm Der}[E_{(1)},\ldots,E_{(\N)}] = \; {\rm det}\left[\sum_i \alpha_{(i)} E_{(i)}\right]R\left[\sum_i \alpha_{(i)} E_{(i)}\right],
\ee
which can be obtained by performing the field re-definition \eqref{fieldredef} on the \EH{} term
\be
{\cal L}_{\rm EH}[E_{(1)}] = \; {\rm det}\left[E_{(1)}\right] R\left[E_{(1)}\right].
\ee
Two specific cases will be of particular interest. First the case, where we have a single dynamical vielbein $E_{(1)}$ and a non-dynamical reference metric (which is a linear superposition of arbitrary non-dynamical vielbeins). Such a term provides a new candidate kinetic interaction for $E_{(1)}$. Let us consider the case where the non-dynamical reference vielbein is flat and we therefore have
\be
\hat{\cal L}_{\rm Kin \; I}[E_{(1)}] = \; {\rm det}\left[\alpha_{(1)} E_{(1)} + \alpha_{(2)} \delta \right]R\left[\alpha_{(1)} E_{(1)} + \alpha_{(2)} \delta \right].
\ee
In fact, we can simplify this even more and consider the equivalent interaction (equivalent up to overall constants of proportionality and a rescaling of the field $E_{(1)}$ -- a re-scaled Planck mass for $E_{(1)}$ if you will)
\be
{\cal L}_{\rm Kin \; I}[E_{(1)}] = \; {\rm det}\left[E_{(1)} + \delta \right]R\left[E_{(1)} + \delta \right].
\ee
Secondly we will be interested in the Bigravity version of \eqref{Der} as a minimal example of the new derivative interactions involving several dynamical fields
\be
{\cal L}_{\rm Der \; II}[E_{(1)},E_{(2)}] = \; {\rm det}\left[E_{(1)} + E_{(2)} \right] R\left[E_{(1)} + E_{(2)} \right],
\ee
where we have again absorbed the constants $\alpha_{(i)}$ for simplicity. Theory graphs depicting the different kinetic and derivative interactions are shown in figure \ref{fig-kinetic}.

\subsection{Kinetic terms}

The new candidate kinetic term for a single spin-2 field $E_{(1)}$ proposed above is
\be \label{KinI}
{\cal L}_{\rm Kin \; I}[E_{(1)}] = \; {\rm det}\left[E_{(1)} + \delta \right]R\left[E_{(1)} + \delta \right].
\ee
Note that, by itself this term is equivalent to the \EH{} term (since they are related by linear field re-definitions and there are no other terms in the action yet) and hence trivially ghost-free. Two observations can already be made at this point. ${\cal L}_{\rm Kin \; I}$ is invariant under local diffeomorphism symmetries and Lorentz transformations \eqref{diff} acting on $E_{(1)} + \delta$ ({\it not} on $E_{(1)}$). Again this is a straightforward consequence of the fact that ${\cal L}_{\rm Kin \; I}$ is related to the \EH{} term via a linear field redefinition. Secondly we can perturb ${\cal L}_{\rm Kin \; I}$ just as we did for the \EH{} term \eqref{EH2},\eqref{EH3},etc. Setting $E_{(1)} = \delta + h$ we then find
\be
{\cal L}_{\rm EH}^{(n)}[h] = 2^{n-2} {\cal L}_{\rm Kin \; I}^{(n)}[h].
\ee
In other words, up to an overall constant of proportionality these two kinetic interactions are the same up to a rescaling in $h$, $h \to \hat h = h/2$. This is of course not surprising, since e.g. $R\left[E_{(1)}+\delta\right] = R\left[2\delta + h\right] \propto R\left[\delta + \hat h\right]$.

\subsection{Derivative interactions}

If more than one dynamical field is present, \eqref{Der} describes new derivative interactions between these fields. For simplicity let us focus on the two field case
\be \label{derII}
{\cal L}_{\rm Der \; II}[E_{(1)},E_{(2)}] = \; {\rm det}\left[E_{(1)} + E_{(2)} \right] R\left[E_{(1)} + E_{(2)} \right].
\ee
Since we are working in the vielbein formulation, all conclusions will straightforwardly generalise to the case of several fields. \eqref{derII} is invariant under local diffeomorphism symmetries and Lorentz transformations \eqref{diff} acting on $E_{(1)} + E_{(2)}$, i.e. under the diagonal subgroup of local diffeomorphism symmetries and Lorentz transformations acting on $E_{(1)}$ and $E_{(2)}$ separately. This straightforwardly generalises to the same symmetries acting on $\sum_i \alpha_{(i)} E_{(i)}$ for \eqref{Der}. In other words, a term such as \eqref{derII} or \eqref{Der} comes with one copy of generalised coordinate invariance $GC_{(i)}$ (cf. \cite{ArkaniHamed:2001ca,ArkaniHamed:2002sp,Noller:2013yja}) linked to its `argument' $(\sum_i \alpha_{(i)} E_{(i)})$. Also, just as for the kinetic term above, by itself ${\cal L}_{\rm Der \; II}$ describes the same physics as the \EH{} term (since it is the \EH{} term with a linear field re-definition) and is consequently trivially ghost-free and consistent. 

It is worth pointing out that ${\cal L}_{\rm Der \; II}$ by itself still only describes the propagating \dof of a single massless spin-2 field, precisely because it is related to the \EH via a linear field redefinition. It is only when other interactions are added to the action that more of the \dof contained in $E_{(1)}$ and $E_{(2)}$ become dynamical -- see below. It may be instructive to consider a quick scalar field example to illustrate this point. We begin with a single massless scalar field $\phi$ with action
\be
{\cal S}_{\phi} = \int d^D x -\frac{1}{2}\pa_\mu \phi \pa^\mu \phi
\ee
and perform the linear field re-definition $\phi \to \phi + \pi$. The resulting action
\be
{\cal S}_{\phi + \pi} = \int d^D x - \tfrac{1}{2} \partial_{a}\phi \partial^{a}\phi -  \partial_{a}\pi \partial^{a}\phi -  \tfrac{1}{2} \partial_{a}\pi \partial^{a}\pi,
\ee
analogous to \eqref{derII}, describes the \dof{} of a single massless scalar and nothing else, despite of the appearance of two fields in the action. However, if we were to add extra interactions for $\phi$ and/or $\pi$, extra propagating \dof{} can be introduced.

For completeness let us repeat our perturbative analysis, this time for ${\cal L}_{\rm Der \; II}[E_{(1)},E_{(2)}]$. We perturb the two vielbeins via
\begin{align}
E_{(1)}{}_{\mu}^{\ A} &= \delta_{\mu}^{\ A} + h_{\mu}^{\ A} &E_{(2)}{}_{\mu}^{\ A} &= \delta_{\mu}^{\ A} + l_{\mu}^{\ A}
\end{align}
and find at quadratic order in the fields
\bea \label{derpert}
{\cal L}_{\rm Der \; II}^{(2)} &=& 2 h^{ab}{}_{,a} l^{c}{}_{c}{}_{,b} -  h^{c}{}_{c}{}_{,b} h^{a}{}_{a}{}^{,b} - 2 l^{c}{}_{c}{}_{,b} h^{a}{}_{a}{}^{,b} -  l^{c}{}_{c}{}_{,b} l^{a}{}_{a}{}^{,b} + h^{ab}{}_{,b} h_{a}{}^{c}{}_{,c}  + 2 h^{ab}{}_{,a} h_{b}{}^{c}{}_{,c} \nonumber \\ 
&& + 2 h^{a}{}_{a}{}^{,b} h^{c}{}_{b}{}_{,c} + 2 h^{ab}{}_{,b} l_{a}{}^{c}{}_{,c} + l^{ab}{}_{,b} l_{a}{}^{c}{}_{,c} + 2 h^{ab}{}_{,a} l_{b}{}^{c}{}_{,c} + 2 l^{ab}{}_{,a} l_{b}{}^{c}{}_{,c} + 2 h^{ab}{}_{,b} l^{c}{}_{a}{}_{,c} \nonumber \\ 
&& + 2 h^{a}{}_{a}{}^{,b} l^{c}{}_{b}{}_{,c} + 2 l^{a}{}_{a}{}^{,b} l^{c}{}_{b}{}_{,c} -  h_{bc}{}_{,a} h^{ab}{}^{,c}  -  \tfrac{5}{2} h_{cb}{}_{,a} h^{ab}{}^{,c} -  l_{bc}{}_{,a} h^{ab}{}^{,c} - 5 l_{cb}{}_{,a} h^{ab}{}^{,c} \nonumber \\ 
&& -  \tfrac{3}{2} h_{ac}{}_{,b} h^{ab}{}^{,c} - 3 l_{ac}{}_{,b} h^{ab}{}^{,c} -  l_{ca}{}_{,b} h^{ab}{}^{,c}  + \tfrac{5}{2} h_{ab}{}_{,c} h^{ab}{}^{,c} -  \tfrac{3}{2} h_{ba}{}_{,c} h^{ab}{}^{,c} + 5 l_{ab}{}_{,c} h^{ab}{}^{,c} \nonumber \\ 
&& - 3 l_{ba}{}_{,c} h^{ab}{}^{,c} -  l_{bc}{}_{,a} l^{ab}{}^{,c}  -  \tfrac{5}{2} l_{cb}{}_{,a} l^{ab}{}^{,c} -  \tfrac{3}{2} l_{ac}{}_{,b} l^{ab}{}^{,c} + \tfrac{5}{2} l_{ab}{}_{,c} l^{ab}{}^{,c} -  \tfrac{3}{2} l_{ba}{}_{,c} l^{ab}{}^{,c}.
\eea
We will refrain from showing cubic order (which contains $\sim$ 500 terms) and higher here.

\begin{figure}[tp]
\centering
\begin{tikzpicture}[-,>=stealth',shorten >=0pt,auto,node distance=2cm,
  thick,main node/.style={circle,fill=blue!30,draw,font=\sffamily\large\bfseries},arrow line/.style={thick,-},barrow line/.style={thick,->},no node/.style={plain},rect node/.style={rectangle,fill=blue!10,draw,font=\sffamily\large\bfseries},red node/.style={rectangle,fill=red!10,draw,font=\sffamily\large\bfseries},green node/.style={circle,fill=green!20,draw,font=\sffamily\large\bfseries},yellow node/.style={rectangle,fill=yellow!20,draw,font=\sffamily\large\bfseries}]

  \node[fill=white!30,line width=0.5mm,draw=black,double](199){}; 
  \node[fill=white!30,line width=0.5mm,draw=black,double](200)[right of=199]{};      
   \path (200) edge[loop below,-,dashed] node[draw=none,left=1pt]{} (300); 
   \node[draw=none,fill=none](84)[right=0.5cm of 199]{};

 \node[fill=white!30,thin,draw=black](100)[right=2.5cm of 200]{};
  \node[fill=white!30,thin,draw=black] (101) [right of=100] {};
 
 \node[fill=white!30,line width=0.5mm,draw=black,double](103)[right=2.5cm of 101]{};
  \node[fill=white!30,line width=0.5mm,draw=black,double] (104) [right of=103] {};

         \node[draw=none,fill=none](83)[right=0.5cm of 100]{};
         \node[draw=none,fill=none](85)[right=0.5cm of 103]{};

\draw[-,line width=.5mm] (100) to (101);
\draw[-,line width=.5mm] (103) to (104);

   \node[draw=none,fill=none,rectangle](92)[below of=84]{\rm \normalsize (a) {\it EH} terms};
   
   \node[draw=none,fill=none,rectangle](93)[below of=83]{\rm \normalsize (b) ${\cal L}_{\rm Der \; II}$};
   
   \node[draw=none,fill=none,rectangle](94)[below of=85]{\rm \normalsize (c) Theory \eqref{2EHD}};
\end{tikzpicture}
\caption{Theory graphs for kinetic and derivative interactions: These are denoted by double lines around nodes or thick lines connecting nodes. a) Two fields, each equipped with an \EH{} term as denoted by the double circles around the node (the second node also has a cosmological constant interaction as indicated by the loop). b) A derivative interaction linking two fields of the type \eqref{derII}. If one of the fields is frozen to become non-dynamical, this is the interaction \eqref{KinI}. In terms of theory graphs these two cases look the same (as long as there are no further terms in the theory that break the degeneracy). c) The theory \eqref{2EHD}, which combines ${\cal L}_{\rm Der \; II}$ with two \EH{} terms.} \label{fig-kinetic}
\end{figure}
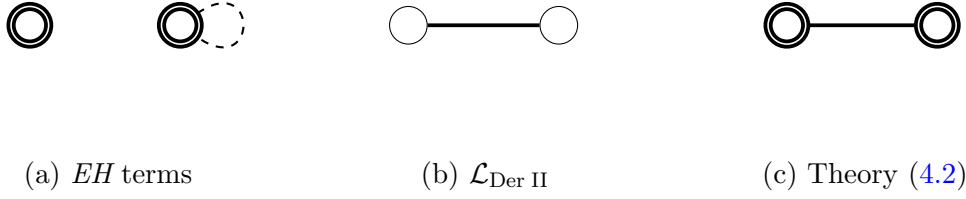

\subsection{Adding self-interactions: `Potential' and `Kinetic' frames}

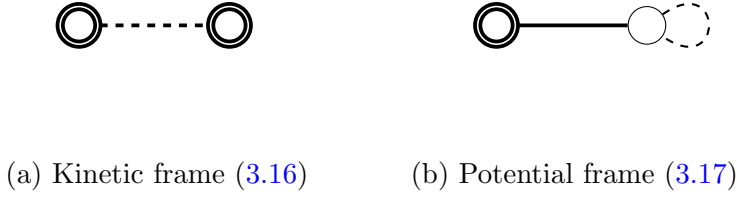
\begin{figure}[tp]
\centering
\begin{tikzpicture}[-,>=stealth',shorten >=0pt,auto,node distance=2cm,
  thick,main node/.style={circle,fill=blue!30,draw,font=\sffamily\large\bfseries},arrow line/.style={thick,-},barrow line/.style={thick,->},no node/.style={plain},rect node/.style={rectangle,fill=blue!10,draw,font=\sffamily\large\bfseries},red node/.style={rectangle,fill=red!10,draw,font=\sffamily\large\bfseries},green node/.style={circle,fill=green!20,draw,font=\sffamily\large\bfseries},yellow node/.style={rectangle,fill=yellow!20,draw,font=\sffamily\large\bfseries}]

 \node[fill=white!30,line width=0.5mm,draw=black,double](100)[right=3.5cm of 200]{};
  \node[fill=white!30,line width=0.5mm,draw=black,double] (101) [right of=100] {};
 
 \node[fill=white!30,line width=0.5mm,draw=black,double](103)[right=3cm of 101]{};
  \node[fill=white!30,thin,draw=black] (104) [right of=103] {}; 
  
   \path (104) edge[loop below,-,dashed] node[draw=none,left=1pt]{} (300);

         \node[draw=none,fill=none](83)[right=0.5cm of 100]{};
         \node[draw=none,fill=none](85)[right=0.5cm of 103]{};

\draw[-,line width=.5mm,dashed] (100) to (101);
\draw[-,line width=.5mm] (103) to (104);

   \node[draw=none,fill=none,rectangle](93)[below of=83]{\rm \normalsize (a) Kinetic frame \eqref{kinetic2}};
   
   \node[draw=none,fill=none,rectangle](94)[below of=85]{\rm \normalsize (b) Potential frame \eqref{potential2}};
   
\end{tikzpicture}
\caption{Here we show a particular bigravity theory in its kinetic frame version \eqref{kinetic2} and in its potential frame version \eqref{potential2}. The two cases are related by linear field re-definitions of the vielbeins, involve \EH{} terms, the new derivative intraction ${\cal L}_{\rm Der \; II}$ and potential self-interactions of the bigravity and cosmological constant type.} \label{fig-kinpotframe}
\end{figure}

Let us now move on and consider adding self-interactions for the spin-2 fields to the kinetic and derivative terms discussed above (which by themselves described the exact same physics as isolated \EH{} terms). 
\\

{\bf A single field:} The complete set of GR interactions consists of the two Lovelock invariants in 4D: The \EH{} term plus a cosmological constant potential self-interaction term 
\bea \label{GR}
{\cal S}_{\rm GR} &=& \int d^D x {\rm det}[E] R[E] + \int d^D x {\rm det}[E] \Lambda\nn\\
 &=& \int d^D x {\cal L}_{\rm EH}[E] + \int d^D x {\rm det}[E] \Lambda,
\eea
where $\Lambda$ is a (cosmological) constant. If we linearly redefine fields throughout this whole action, we have obviously not changed anything. However, if we only do this with the field re-definition \eqref{fieldredef} for one of the terms (i.e. we do {\it not} perform a linear field redefinition at the level of the {\it full} action), then we obtain a different action. Consider one specific such case
\bea 
{\cal S}_1 &=& \int d^D x {\rm det}[E+\delta] R[E+\delta] + \int d^D x {\rm det}[E] \Lambda \nn\\
&=& \int d^D x {\cal L}_{\rm Kin \; I}[E] + \int d^D x {\rm det}[E] \Lambda.
\label{KinI-predef}
\eea
Upon a linear field re-definition $E \to \tilde E - \delta$ (now at the level of the full action, so leading to an equivalent theory to ${\cal S}_1$) this becomes
\bea 
\tilde {\cal S}_1 &=& \int d^D x {\rm det}[\tilde E] R[\tilde E] + \int d^D x {\rm det}[\tilde E - \delta] \Lambda \nn\\
&=& \int d^D x {\cal L}_{\rm EH}[\tilde E] + \int d^D x {\rm det}[\tilde E - \delta] \Lambda.
\label{Kini-redef}
\eea
This is a fully ghost-free massive gravity theory with a particular superposition of ghost-free massive gravity interaction terms of the form discussed above \eqref{detform}.\footnote{Throughout this section we use this form of the potential self-interactions for simplicity, but do note that we could just as well consider e.g. \eqref{Kini-redef} with a general self-interaction of the form \eqref{potential}. In that case the dual theory \eqref{KinI-predef} would not just have cosmological constant-type self-interactions, but more general potential interactions (also of the form \eqref{potential} but with different coefficients) as well.} It has a strong coupling scale $\Lambda_3$, where the associated mass scale $m$ is given via $\Lambda \equiv m^2 M_{\rm Pl}^2$.\footnote{Note that we have set $M_{\rm Pl}^2$ to unity in \eqref{Kini-redef} and \eqref{KinI-predef} as well as ignored overall multiplicative factors.} So in particular this means that \eqref{KinI-predef} is a fully ghost-free theory of a spin-2 field with the new kinetic term ${\cal L}_{\rm Kin \; I}[E]$. The elaborate structure of the massive gravity potential interactions has been traded for the new kinetic term here. We shall refer to an action like \eqref{KinI-predef} as the `potential frame' version of the theory (with maximally simple -- in this case cosmological constant type -- self-interactions, but complicated kinetic interactions), whereas \eqref{Kini-redef} is the `kinetic frame' version of the theory (where kinetic interactions are maximally simple and of \EH{} form).\footnote{Note that, for a fully general spin-2 field theory, one is of course not guaranteed to be able to find a `frame' where all kinetic terms are \EH{} or all the self-interactions are of the cosmological constant type, although this is the case for theories that are valid, ghost-free \eft s all the way to the Planck scale (cf. \cite{deRham:2013tfa}). Even in cases where the `kinetic frame' version only contains \EH{} terms, the potential frame version of the theory will not in general just be a superposition of cosmological constant type terms for the individual fields. This is only the case if the potential self-interactions in the kinetic frame are of the form ${\rm det}\left[\sum_i \alpha_{(i)}E_{(i)}\right]$.} 

Actions \eqref{KinI-predef} and \eqref{Kini-redef} are no longer diffeomorphism invariant, in an interesting way. In \eqref{KinI-predef} (the kinetic frame) the kinetic term is invariant under diffeomorphisms acting on $E+\delta$, the cosmological constant term under diffeomorphisms acting on $E$. In this picture, where $E$ is the fundamental \dof{}, the kinetic term ${\cal L}_{\rm Kin \; I}[E]$ breaks diffeomorphism invariance and hence leads to the propagation of additional \dof. The resulting 5 \dof{} describe a single massive graviton. The dual picture in the potential frame \eqref{Kini-redef} describes a theory where diffeomorphisms for $E$ are broken by the known ghost-free massive gravity interactions. The propagating 5 \dof{} of a massive graviton here arise as a consequence of the symmetry breaking potential type interactions.\footnote{After this paper was published on the arXiv, we became aware of appendix A in \cite{Deffayet:2012nr}, where a construction analogous to our ``kinetic'' and ``potential'' frames is discussed for a single massive graviton. We thank an anonymous referee for pointing this out.} 
\\

{\bf Two fields:} Let us do the same for the Bigravity case (in the vielbein language the general argument for $\N$ vielbeins then straightforwardly follows). We pick a simple, consistent and fully ghost-free bigravity action 
\bea
{\cal S}_2 &=& \int d^D x {\rm det}[E_{(1)}]R[E_{(1)}] + \int d^D x {\rm det}[E_{(2)}]R[E_{(2)}] \nn\\ &+& \int d^D x {\rm det}\left[E_{(1)} + E_{(2)}\right] \Lambda.\nn\\
 &=& \int d^D x {\cal L}_{\rm EH}[E_{(1)}] + \int d^D x {\cal L}_{\rm EH}[E_{(2)}] + \int d^D x {\rm det}\left[E_{(1)} + E_{(2)}\right] \Lambda.\label{kinetic2}
\eea
Such an action is again in the `kinetic' frame, where kinetic terms take on the simple \EH -form and potential self-interactions are of the ghost-free multi-gravity type. The field re-definition $E_{(1)} \to E_{(1)} - E_{(2)}$ then maps this to the potential frame (where potential interactions are simple cosmological constants, but kinetic interactions take on the new form) and which has an action
\bea
\tilde {\cal S}_2 &=& \int d^D x {\rm det}[E_{(1)}- E_{(2)}]R[E_{(1)}- E_{(2)}] + \int d^D x {\rm det}[E_{(2)}]R[E_{(2)}] \nn\\ &+& \int d^D x {\rm det}\left[E_{(1)}\right]\Lambda \nn\\
&=& \int d^D x {\cal L}_{\rm Der \; II}[E_{(1)},-E_{(2)}] + \int d^D x {\cal L}_{\rm EH}[E_{(2)}] + \int d^D x {\rm det}\left[E_{(1)}\right]\Lambda.\label{potential2}
\eea
The potential frame version of the theory therefore describes a fully ghost-free set of interactions involving ${\cal L}_{\rm Der \; II}$. Theory graphs depicting these Kinetic and Potential frame examples are shown in figure \ref{fig-kinpotframe}.
\eqref{kinetic2} has a kinetic sector invariant under diffeomorphisms acting on $E_{(1)}$ and $E_{(2)}$ separately, where the potential interactions break this down to their diagonal subgroup and lead to a total of $2+5$ propagating \dof. In contrast \eqref{potential2} has a potential sector invariant under diffeomorphisms acting on $E_{(1)}$ and a kinetic sector invariant under diffeomorphisms acting on $E_{(1)}-E_{(2)}$ and $E_{(2)}$. So, in terms of diffeomorphisms acting on $E_{(1)}$ and $E_{(2)}$, the ${\cal L}_{\rm Der \; II}$ derivative interaction breaks the overall diffeomorphism symmetry of the theory down to their diagonal subgroup (for details of the symmetry breaking pattern we again refer to \cite{Hinterbichler:2012cn,Noller:2013yja}). Finally, just as discussed above, the qualitative conclusions drawn here would not change had we used a more general potential self-interaction to start with in \eqref{kinetic2}. 
\\

{\bf Several fields:} Analogous examples for generic Multi-Gravity theory with $\N$ fields are easily computed.\footnote{For a precise classification of how many and which \dof{} propagate in a general multi-gravity theory see \cite{Noller:2013yja}.} For example the fully ghost-free (kinetic frame) theory
\bea \label{multiex}
{\cal S}_3 &=& \sum_i^{\N} \int d^Dx {\rm det}[E_{(i)}] R\left[E_{(i)}\right] + \int d^D x {\rm det}\left[\sum_i^{\N} \alpha_{(i)} E_{(i)}\right]\Lambda
\eea
under the linear field re-definition $\alpha_{(1)} E_{(1)} \to \alpha_{(1)} E_{(1)} - \sum_{i=2}^{\N} \alpha_{(i)} E_{(i)}$ maps to
\bea 
\tilde{\cal S}_3 &=& \int d^Dx {\rm det}\left[E_{(1)} + \frac{1}{\alpha_{(1)}}\sum_{i=2}^{\N}\alpha_{(i)}E_{(i)} \right] R\left[E_{(1)} + \frac{1}{\alpha_{(1)}}\sum_{i=2}^{\N}\alpha_{(i)}E_{(i)} \right] \nn \\ &+&
\sum_{i=2}^{\N} \int d^Dx {\rm det}[E_{(i)}] R\left[E_{(i)}\right] + \int d^D x {\rm det}\left[\alpha_{(1)} E_{(1)}\right]\Lambda. \label{multipost}
\eea
Again \eqref{multipost} describes a consistent theory of $\N$ interacting spin-2 fields, where the breaking of several diffeomorphism invariances and the associated additional propagating degrees of freedom are now encoded in the derivative ${\cal L}_{\rm Der}$ terms and takes place just as discussed above.
\\

Let us finish this subsection by emphasizing that the field re-definition trick used to generate the new kinetic interactions and map between frames is not a trivial procedure. It relies on spotting the right set of field re-definitions \eqref{fieldredef} that leave the form of the ghost-free potential interactions unchanged. Taking some other linear field re-definition and selectively applying it to some of the terms in a consistent spin-2 field theory (like we did in going from \eqref{GR} to \eqref{KinI-predef}) does {\it not} generate another consistent theory. For example, taking a theory with the \EH{} term in metric form, $\sqrt{-g}R[g]$, plus a cosmological constant term for $g$ and then replacing $g \to g+\eta$ only in the kinetic term does {\it not} result in a ghost-free theory. In other words,  $\int d^Dx \sqrt{-(g+\eta)}R[g+\eta] + \sqrt{-g}\Lambda$ is not a ghost-free theory of a spin-2 field.

\subsection{Adding a coupling to matter: `Einstein' and `Jordan' frames}

\begin{figure}[tp]
\centering
\begin{tikzpicture}[-,>=stealth',shorten >=0pt,auto,node distance=2cm,
  thick,main node/.style={circle,fill=blue!30,draw,font=\sffamily\large\bfseries},arrow line/.style={thick,-},barrow line/.style={thick,->},no node/.style={plain},rect node/.style={rectangle,fill=blue!10,draw,font=\sffamily\large\bfseries},red node/.style={rectangle,fill=red!10,draw,font=\sffamily\large\bfseries},green node/.style={circle,fill=green!20,draw,font=\sffamily\large\bfseries},yellow node/.style={rectangle,fill=yellow!20,draw,font=\sffamily\large\bfseries}]

 \node[fill=blue!30,line width=0.5mm,draw=black,double](100)[right=3.5cm of 200]{};
  \node[fill=blue!30,line width=0.5mm,draw=black,double] (101) [right of=100] {};
 
 \node[fill=white!30,line width=0.5mm,draw=black,double](103)[right=3cm of 101]{};
  \node[fill=blue!30,thin,draw=black] (104) [right of=103] {};

         \node[draw=none,fill=none](83)[right=0.5cm of 100]{};
         \node[draw=none,fill=none](85)[right=0.5cm of 103]{};

\draw[-,line width=.5mm] (103) to (104);

   \node[draw=none,fill=none,rectangle](93)[below of=83]{\rm \normalsize (a) Einstein frame \eqref{S4}};
   
   \node[draw=none,fill=none,rectangle](94)[below of=85]{\rm \normalsize (b) Jordan frame \eqref{S42}};
   
\end{tikzpicture}
\caption{Here we show a particular bigravity theory in its Einstein frame version \eqref{S4} and in its Jordan frame version \eqref{S42}. The two cases are related by linear field re-definitions of the vielbeins, involve \EH{} terms, the new derivative interaction ${\cal L}_{\rm Der \; II}$ and minimal GR-like matter couplings as well as the more general new matter couplings \eqref{matviel}.}  \label{fig-EinJorframe}
\end{figure}
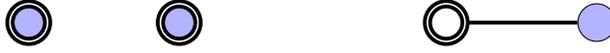

Having considered `kinetic' and `potential' frames above, including a coupling to matter and working out the `Einstein' and `Jordan' frame pictures for Massive-, Bi- and Multi-Gravity now becomes straightforward. The general theories for spin-2 fields coupled to matter discussed in \ref{subsec-matter} above have the form
\be \label{matterEJ}
{\cal S}_3 = \sum_i^{\N} \int d^D x {\rm det}[E_{(i)}]R[E_{(i)}] + \int d^D x {\rm det}\left[\sum_i^{\N} \alpha_{(i)} E_{(i)}\right] {\cal L}\left[\Phi_i, \sum_i^{\N} \alpha_{(i)} E_{(i)} \right].
\ee
The standard minimal coupling we know from GR is contained as a special case with some $\alpha_{i} = 1$ and $\alpha_{j}=0$ for all $j \ne i$. Again, as discussed in \ref{subsec-matter}, \eqref{matterEJ} has a ghost at the scale $\Lambda_g$, but is a valid and consistent \eft{} at all scales below $\Lambda_g$ or $\Lambda_c$, whichever is lower. 

For simplicity let us again focus on a particular Bigravity case and ignore additional potential self-interactions.\footnote{Omitting potential self-interactions will not change any of the conclusions of this section, since these preserve their overall form under linear field-redefinitions of the vielbeins. Also note that the massive gravity case can be obtained by setting $E_{(2)} = \delta$ in the bigravity action \eqref{S4}.} The particular action we consider is
\bea
{\cal S}_4 &=& \int d^D x {\rm det}[E_{(1)}]R[E_{(1)}] + \int d^D x {\rm det}[E_{(2)}]R[E_{(2)}] \nn\\ &+& \int d^D x {\rm det}\left[E_{(1)} + E_{(2)}\right] {\cal L}\left[\Phi_i,E_{(1)} + E_{(2)} \right].
\label{S4}
\eea
Up to the scale of the ghost $\Lambda_g$ (and at the very least up to $\Lambda_3$) this is a fully consistent \eft. ${\cal S}_4$ is an example of an `Einstein' frame action, where kinetic interactions are simply a superposition of the standard \EH{} terms for individual spin-2 fields. The field re-definition $E_{(1)} \to E_{(1)} - E_{(2)}$ then maps \eqref{S4} to a frame, which we may call the Jordan frame and which has an action
\bea
\tilde{\cal S}_4 &=& \int d^D x {\rm det}[E_{(1)}- E_{(2)}]R[E_{(1)}- E_{(2)}] + \int d^D x {\rm det}[E_{(2)}]R[E_{(2)}] \nn\\ &+& \int d^D x {\rm det}\left[E_{(1)}\right] {\cal L}\left[\Phi_i,E_{(1)}\right].\label{S42}
\eea
By construction \eqref{S42} also has to have a ghost at the scale $\Lambda_g$, but is a healthy \eft{} below that scale. \eqref{S42} is consequently an example of a theory that includes the new kinetic interaction ${\cal L}_{\rm Der \; II}[E_{(1)},-E_{(2)}]$ and in addition an \EH{} term and a minimally coupled matter sector. All these terms are ghost-free individually, but still give rise to a ghost when superposed. This is an explicit example of the point discussed above, that superposing fully consistent actions is {\it not} guaranteed to yield a consistent theory for all scales. It also alerts us to the fact that the new kinetic and derivative interactions proposed here cannot simply be added to any known consistent action without adding instabilities at some scale, making our findings fully consistent with the `no-go' theorem of \cite{deRham:2013tfa}. In \cite{kinconstraint} we will investigate the scale of the associated ghost $\Lambda_g$ -- the `Jordan frame' picture \eqref{S42} should prove very useful in this task, decoupling the non-trivial part of the action from the sector that involves a direct coupling to all matter species. We again emphasize that below their cutoff scale, theories \eqref{S4} and \eqref{S42} are perfectly healthy \eft s -- if $\Lambda_g \gg \Lambda_3$ these will have an interesting non-linear regime with non-trivial matter couplings and/or kinetic and derivative interactions. Theory graphs depicting these Einstein and Jordan frame examples are shown in figure \ref{fig-EinJorframe}.

As before it is straightforward to extend the above to a generic Multi-Gravity case. For example consider the `Einstein frame' action
\be \label{multiexM}
{\cal S}_5 = \sum_i^{\N} \int d^D x {\rm det}[E_{(i)}]R[E_{(i)}] + \int d^D x {\rm det}\left[\sum_i^{\N} \alpha_{(i)} E_{(i)}\right] {\cal L}\left[\Phi_i, \sum_i^{\N} \alpha_{(i)} E_{(i)} \right].
\ee
Under the linear field re-definition $\alpha_{(1)} E_{(1)} \to \alpha_{(1)} E_{(1)} - \sum_{i=2}^{\N} \alpha_{(i)} E_{(i)}$ this action maps to the `Jordan frame' action
\bea 
\tilde{\cal S}_5 &=& \int d^Dx {\rm det}\left[E_{(1)} + \frac{1}{\alpha_{(1)}}\sum_{i=2}^{\N}\alpha_{(i)}E_{(i)} \right] R\left[E_{(1)} + \frac{1}{\alpha_{(1)}}\sum_{i=2}^{\N}\alpha_{(i)}E_{(i)} \right] \nn \\ &+&
\sum_{i=2}^{\N} \int d^Dx {\rm det}[E_{(i)}] R\left[E_{(i)}\right] + \int d^D x {\rm det}\left[\alpha_{(1)} E_{(1)}\right]{\cal L}\left[\Phi_i, \alpha_{(1)} E_{(1)} \right]. \label{multipostM}
\eea

\subsection{Other candidate kinetic and derivative interactions}

In the above we have constructed new kinetic and derivative interactions, which can be included in the construction of completely ghost-free pure spin-2 theories such as \eqref{KinI-predef}, \eqref{kinetic2} and \eqref{multipost} or in the construction of spin-2 theories coupled to matter, which are valid \eft s up to the scale $\Lambda_g$ (or, in the best case: $\Lambda_c$), such as \eqref{S42} and \eqref{multipostM}. Other candidate kinetic terms, different from the \EH{} term, exist \cite{deRham:2013tfa}. These terms are ${\cal S}^{4d}_{{\cal K}G}$ and ${\cal S}^{4d}_{{\cal KK^{\star}}R}$ and can be obtained by considering a discretised 5D Gauss-Bonnet Lagrangian. ${\cal S}^{4d}_{{\cal K}G}$ can also be seen as a non-linear completion of the linearised \EH{} term, which, at non-linear orders, is different from the \EH{} term. In a similar vein, ${\cal S}^{4d}_{{\cal KK^{\star}}R}$ can be seen as a non-linear completion of the candidate linear kinetic term ${\cal L}_3^{(der)}$ , proposed by \cite{Hinterbichler:2013eza} (also see \cite{Folkerts:2011ev,Kimura:2013ika,Folkerts:2013mra,Gao:2014jja}). As shown by \cite{deRham:2013tfa}, both ${\cal S}^{4d}_{{\cal K}G}$ and ${\cal S}^{4d}_{{\cal KK^{\star}}R}$ are ghost-free in the decoupling limit (i.e. up to the scale $\Lambda_3$), but inevitably contain a ghost at some larger scale. Also note that, in the metric formulation, \cite{Hassan:2012wr} discuss kinetic terms arising from field re-definitions explicitly separating out the massive modes of the theory. 

${\cal S}^{4d}_{{\cal K}G}$ and ${\cal S}^{4d}_{{\cal KK^{\star}}R}$ are therefore different from the ${\cal L}_{\rm Der}$ terms proposed here, which are completely ghost-free by themselves (because they {\it are} just \EH{} in that case), can be completely ghost-free when self-interaction terms are added as discussed above, but also generically contain a ghost at a scale $\Lambda_g$ (e.g. when a general coupling to matter is added). However, the same logic applies for all of these terms: Even if there is a ghost present in the theory with any of the candidate kinetic and derivative terms at a scale $\Lambda_g$, below that scale the theory can still be a valid and ghost-free \eft{}. If $\Lambda_g \gg \Lambda_3$, in addition there will be interesting non-linear phenomenology (such as Vainshtein screening) described within the regime of validity of the theory. We investigate all the candidate kinetic terms and identify the associated scales $\Lambda_g$ in \cite{kinconstraint}. It would also be interesting to consider Multi-Gravity generalisations of ${\cal S}^{4d}_{{\cal K}G}$ and ${\cal S}^{4d}_{{\cal KK^{\star}}R}$ as candidate derivative interactions involving more than one dynamical spin-2 field.

\section{General spin-2 theories}\label{sec-genspin2}

Having constructed a set of new kinetic interactions and established their relation with `Kinetic' and `Potential'/`Einstein' and `Jordan' frames, in this section we point out some theories involving the new kinetic/derivative terms that merit further investigation and extend the theory graph machinery of \cite{ArkaniHamed:2001ca,ArkaniHamed:2002sp,Hinterbichler:2012cn,Noller:2013yja} in order to have a diagrammatic way of representing theories involving the new matter coupling and kinetic/derivative interactions, which makes the symmetry (and symmetry-breaking) structure of those theories explicit. 

\subsection{Arbitrary superpositions of terms}\label{subsec-arbsuper}

In the previous sections we spotted a set of field re-definitions that leave the form of consistent self-interaction potentials \eqref{potential} and the general matter couplings \eqref{matviel} unchanged. We then used this to propose a new set of kinetic and derivative terms, related to the \EH{} term(s) via the same set of field re-definitions. This had two effects: 1) We established the Kinetic/Potential and Einstein/Jordan frames for Massive, Bi- and Multi-Gravity. 2) In the process we found a set of actions involving the new kinetic and derivative interactions, which were related to known theories via linear field re-definitions and hence either fully ghost-free (in the case of some specific theories of just spin-2 fields) or had a ghost at a scale $\Lambda_g > \Lambda_3$ (in the case of general theories involving a non-trivial matter coupling). 

What happens when we consider actions that involve the new kinetic and derivative interactions, but where the new terms cannot all be turned into \EH{} terms via linear field re-definitions? In other words, having generated particular combinations of the new kinetic and derivative interactions via the trick discussed above, can we now deconstruct those particular combinations and build arbitrary combinations of the new kinetic/derivative terms (and pair them with potential and matter coupling terms), which are consistent?\footnote{As discussed in \ref{subsec-inv} we could have discovered particular superpositions of the set of consistent spin-2 potential interaction terms in an analogous fashion. For the potential terms of course every linear superposition of such terms is consistent -- here we are asking to what extent the same holds for the new kinetic and derivative terms.} Generically we should expect these theories to have a ghost (as indicated by the matter coupling considerations above), but at what scales does the ghost enter? Some simple examples of actions where the ${\cal L}_{\rm Der}$ terms cannot be removed with field re-definitions are
\bea 
{\cal S}_6 &=& M_{(1)}^2 \int d^D x {\rm det}[E+\delta] R[E+\delta] + M_{(2)}^2 \int d^D x {\rm det}[E] R[E], \\
{\cal S}_7 &=& M_{(1)}^2 \int d^D x {\rm det}[E_{(1)}]R[E_{(1)}] + M_{(2)}^2 \int d^D x {\rm det}[E_{(1)}+ E_{(2)}]R[E_{(1)}+ E_{(2)}] \nn\\ &+& M_{(3)}^2 \int d^D x {\rm det}[E_{(2)}]R[E_{(2)}], \label{2EHD} \\
{\cal S}_8 &=& \int d^D x {\rm det}[E_{(1)}]R[E_{(1)}] + \int d^D x {\rm det}[E_{(1)}- E_{(2)}]R[E_{(1)}- E_{(2)}] \nn\\ &+& \int d^D x {\rm det}[E_{(1)} + E_{(2)}]R[E_{(1)}+ E_{(2)}] + \int d^D x {\rm det}\left[E_{(1)} + E_{(2)}\right] {\cal L}\left[\Phi_i,E_{(1)} + E_{(2)} \right].\nn\\
\eea

We discuss these examples and others in \cite{kinconstraint} (showing that the above examples ${\cal S}_6, {\cal S}_7, {\cal S}_8 $ generically have ghosts and discussing at what scale the ghost enters). Here we just wish to emphasize a few final observations about the new kinetic and derivative terms. From \cite{Boulanger:2000rq} we know that there can be no fully consistent theory of interacting massless spin-2 fields (where the fields interact with one another), but generic combinations of the new kinetic and derivative terms proposed here and \EH{} terms will break diffeomorphism symmetries and lead to the propagation of massive spin-2 \dof, so the construction here generically evades this no-go theorem. The terms proposed here are equivalent to \EH{} when considered by themselves, but generic combinations of the new terms with \EH{} terms, potential self-interactions and general matter couplings will destroy this equivalence and generically introduce a ghost. This explains the consistency of the new terms with the kinetic no-go theorem of \cite{deRham:2013tfa} (since the no-go theorem of \cite{deRham:2013tfa} applies to new kinetic terms which are ghost-free at all scales up to $M_{Pl}$). Establishing the scale of the ghost $\Lambda_g$ in generic setups will determine what the regime of validity for the resulting \eft s is. In particular if $\Lambda_g \gg \Lambda_3$ the resulting theories will have interesting non-linear physics and hence the potential for screening mechanisms as well.\footnote{If e.g. ${\cal S}_6, {\cal S}_7$ or ${\cal S}_8$ satisfy this criterion, then the derivative interactions proposed in this paper can be used to build interesting models describing distinct physics from models without these new terms. Otherwise the kinetic/potential and Einstein/Jordan frame pictures developed throughout this paper are useful different representations of consistent theories, but all interesting models with the new derivative interactions proposed in this paper would be ultimately dual to models without these interactions. We discuss this in further detail in \cite{kinconstraint}.}

\subsection{Theory graphs}\label{subsec-graph}

\begin{figure}[tp]
\centering
\begin{tikzpicture}[-,>=stealth',shorten >=0pt,auto,node distance=2cm,
  thick,main node/.style={circle,fill=blue!30,draw,font=\sffamily\large\bfseries},arrow line/.style={thick,-},barrow line/.style={thick,->},no node/.style={plain},rect node/.style={rectangle,fill=blue!10,draw,font=\sffamily\large\bfseries},red node/.style={rectangle,fill=red!10,draw,font=\sffamily\large\bfseries},green node/.style={circle,fill=green!20,draw,font=\sffamily\large\bfseries},yellow node/.style={rectangle,fill=yellow!20,draw,font=\sffamily\large\bfseries}]

  \node[fill=blue!30,line width=.5mm,draw=black,double](200){};      
   \path (200) edge[loop below,-,dashed] node[draw=none,left=1pt]{} (300);

 \node[fill=blue!30,line width=.5mm,draw=black,double](100)[right=3.5cm of 200]{};
  \node[fill=white!30,thin,draw=black] (101) [right of=100] {};
 
 \node[fill=blue!30,line width=.5mm,draw=black,double](103)[right=3cm of 101]{};
  \node[fill=white!30,line width=.5mm,draw=black,double] (104) [right of=103] {};

         \node[draw=none,fill=none](83)[right=0.5cm of 100]{};
         \node[draw=none,fill=none](85)[right=0.5cm of 103]{};

\draw[-,line width=.5mm,dashed] (100) to (101);
\draw[-,line width=.5mm,dashed] (103) to (104);

   \node[draw=none,fill=none,rectangle](92)[below of=200]{\rm \normalsize (a) GR};
   
   \node[draw=none,fill=none,rectangle](93)[below of=83]{\rm \normalsize (b) dRGT Massive Gravity};
   
   \node[draw=none,fill=none,rectangle](94)[below of=85]{\rm \normalsize (c) HR Bigravity};
   
\end{tikzpicture}
\caption{Using the theory graph conventions developed throughout this paper, we can now succintly depict the actions for a) GR, with a single spin-2 field endowed with an \EH{} term (the doubly circled node), which is coupled to matter minimally (the node is dark-shaded) and in principle has a cosmological constant term (the dashed loop). b) dRGT Massive Gravity, with a single dynamical metric (the doubly circled node indicating an \EH{} term for this field) with that dynamical metric minimally coupled to matter (a single dark-shaded node) and a potential type interaction (dashed line) linking this metric to a non-dynamical reference metric (the white node without a double-circle). c) Hassan-Rosen Bigravity with two dynamical spin-2 fields, linked by potential type interactions, but with only one field directly (and minimally) coupling to matter.} \label{fig-GRMGHR}
\end{figure}

In light of the new kinetic and derivative interactions proposed here and the matter coupling proposed in \cite{deRham:2014naa,Noller:2014sta}, it is useful to extend the theory graph conventions for Multi-Gravity theories \cite{ArkaniHamed:2001ca,ArkaniHamed:2002sp,Hinterbichler:2012cn,Noller:2013yja} to allow us to depict a general spin-2 theory with all the proposed interaction terms (in the general case this will mean consistent up to some scale $\Lambda_g > \Lambda_3$).\footnote{Note that we have slightly modified the conventions for plotting potential self-interactions from \cite{Noller:2013yja} in order to avoid ambiguities when plotting the new kinetic/derivative and matter coupling interactions.}

Every node in a theory graph corresponds to a spin-2 field/vielbein $E_{(i)}$. In principle each such field comes equipped with its own diffeomorphism symmetry and associated copy of generalised co-ordinate invariance $GC_{(i)}$. Dashed lines connecting nodes denote potential-like interactions involving these fields.\footnote{We emphasize that the theory graphs encode no information about the value of coupling constants in any of the interactions.} A dashed loop around a node consequently corresponds to a cosmological constant term for this field. See figure \ref{fig-potentials} for examples. Nodes encircled by a double boundary have their own \EH{} term and nodes connected by thick lines have one of the new kinetic/derivative interactions discussed in this paper -- see figure \ref{fig-kinetic}. Finally, all dark-shaded nodes are coupling to matter in a way that explicitly upholds the weak equivalence principle as discussed in \ref{subsec-matter} -- see figure \ref{fig-matter}. Figure \ref{fig-GRMGHR} then shows the resulting theory graphs for some familiar examples: GR, dRGT ghost-free massive gravity and Hassan-Rosen Bigravity, fully specifying the kinetic and self-interaction potential for these theories as well as the way they couple to matter. 

\begin{figure}[tp]
\centering
\begin{tikzpicture}[-,>=stealth',shorten >=0pt,auto,node distance=2cm,
  thick,main node/.style={circle,fill=blue!30,draw,font=\sffamily\large\bfseries},arrow line/.style={thick,-},barrow line/.style={thick,->},no node/.style={plain},rect node/.style={rectangle,fill=blue!10,draw,font=\sffamily\large\bfseries},red node/.style={rectangle,fill=red!10,draw,font=\sffamily\large\bfseries},green node/.style={circle,fill=green!20,draw,font=\sffamily\large\bfseries},yellow node/.style={rectangle,fill=yellow!20,draw,font=\sffamily\large\bfseries}]

 \node[fill=blue!30,line width=.5mm,draw=black,double](1001){};
 
 \node[fill=blue!30,thin,draw=black](1031)[right=2cm of 1001]{};
  \node[fill=blue!30,thin,draw=black] (1041) [right of=1031] {};

\draw[-,line width=.5mm] (1031) to (1041);

 \node[draw=none,fill=none](851)[right=0.5cm of 1031]{};

   \node[draw=none,fill=none,rectangle](931)[below of=1001]{\rm \normalsize (a) Theory \eqref{S9}};
   
   \node[draw=none,fill=none,rectangle](941)[below of=851]{\rm \normalsize (b) Dual to (a)};

 \node[fill=blue!30,line width=.5mm,draw=black,double](100)[right=2cm of 1041]{};
 
 \node[fill=blue!30,thin,draw=black](103)[right of=100]{};
  \node[fill=white!30,line width=.5mm,draw=black,double] (104) [right of=103] {};

\draw[-,line width=.5mm,double] (100) to (103);
\draw[-,line width=.5mm,dashed] (103) to (104);

 \node[draw=none,fill=none](85)[right=0.5cm of 103]{};

   \node[draw=none,fill=none,rectangle](93)[below of=103]{\rm \normalsize (a) Theory \eqref{S10}};

\end{tikzpicture}
\caption{(a) and (b) depict the theories \eqref{S9} , which are equivalent upon a linear field re-definition. This theory therefore only propagates the \dof{} of a {\it single} massless spin-2 field (plus the \dof{} of the matter fields of course). This serves to illustrate that it is not always trivial to read off the number of propagating \dof{} from theory graphs. (c) Here we diagrammatically represent \eqref{S10}, a more complex theory combining all the different types of interactions discussed throughout this paper.} \label{fig-final}
\end{figure}
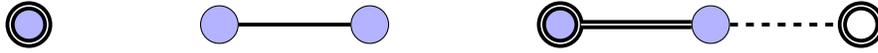

The theory graphs constructed in this way make the symmetry properties of the given theory explicit. Every connected island has one remaining, unbroken copy of diffeomorphism invariance. This is true as long as all the fields involved are dynamical, which corresponds to having the same number of derivative interactions (double circles and thick lines) involving linearly independent combinations of the vielbeins as we have fields in that connected island. If some of the participating fields are non-dynamical, all diffeomorphism invariances are broken -- dRGT massive gravity as depicted in figure \ref{fig-GRMGHR} is an example of such a theory (it has two nodes/fields, but only one kinetic term/double circle). Note, however, that the new matter coupling and kinetic interactions make reading off the number of propagating \dof{} from a theory graph somewhat more complex than before. If a given field $E_{(i)}$ can be completely eliminated from the theory with linear field re-definitions, then it clearly doesn't describe any additional \dof. A concrete example of this is
\bea
{\cal S}_9 &=& \int d^D x {\rm det}[E_{(1)} + E_{(2)}]R[E_{(1)}+ E_{(2)}] + \int d^D x {\rm det}\left[E_{(1)} + E_{(2)}\right] {\cal L}\left[\Phi_i,E_{(1)} + E_{(2)} \right] \nn \\ 
&\to & \int d^D x {\rm det}[E_{(1)}]R[E_{(1)}] + \int d^D x {\rm det}\left[E_{(1)}\right] {\cal L}\left[\Phi_i,E_{(1)} \right], \label{S9}
\eea
where we have performed the linear field re-definition $E_{(1)} \to E_{(1)} - E_{(2)}$ in going to the second line. Note that we could have added a potential self-interaction ${\rm det}[E_{(1)} + E_{(2)}] \Lambda$ and field re-defined it into a cosmological constant for a single field as well. In the theory graph picture, cases analogous to \eqref{S9} correspond to a field's potential (dashed lines), kinetic/derivative (thick lines and doubly circled nodes) and matter coupling (dark shaded nodes) interactions all linking the same set of nodes. This is shown in figure \ref{fig-final}. Figure \ref{fig-final} also illustrates how more complex theories combining the different types of interactions discussed throughout this paper can be depicted by showing the theory graph for the following theory\footnote{Note that the theory graphs only keep track of which fields are linked by interaction terms. Two terms linking the same fields, but differing by their choice of interaction coefficients will look identical -- hence the double-line in figure \ref{fig-final} resulting from the two kinetic interactions ${\rm det}[E_{(1)} + E_{(2)}]R[E_{(1)}+ E_{(2)}]$ and ${\rm det}[E_{(1)} - E_{(2)}]R[E_{(1)} - E_{(2)}]$. Those two terms can of course be combined into a single derivative interaction term, but we here choose to keep the composition in terms of \EH{}-like objects explicit.}
\bea
{\cal S}_9 &=& \int d^D x {\rm det}[E_{(1)}]R[E_{(1)}] + \int d^D x {\rm det}[E_{(1)}- E_{(2)}]R[E_{(1)}- E_{(2)}] \nn\\ &+& \int d^D x {\rm det}[E_{(1)} + E_{(2)}]R[E_{(1)}+ E_{(2)}] + \int d^D x {\rm det}\left[E_{(1)} + E_{(2)}\right] {\cal L}\left[\Phi_i,E_{(1)} + E_{(2)} \right]\nn\\
&+& \int d^D x {\rm det}[E_{(3)}]R[E_{(3)}] + \int d^D x {\rm det}[E_{(3)} + E_{(2)}]R[E_{(3)}+ E_{(2)}].\label{S10}
\eea

\section{Summary and Conclusions}\label{sec-conc}

In this paper we have proposed and discussed a set of new kinetic and derivative interactions for spin-2 fields. These were closely linked to the Einstein and Jordan frame pictures for Massive, Bi- and Multi-Gravity, which we described and which connect the new kinetic/derivative interactions to the matter couplings proposed by \cite{deRham:2014naa,Noller:2014sta}. As such the new kinetic/derivative interactions, when combined with general potential self-interactions and matter coupling terms, are expected to generically inherit a ghost (in several cases outside of the decoupling limit) at a scale $\Lambda_g$ -- we will establish the exact scale $\Lambda_g$ of this ghost and how it relates to the cutoff scale of the theory in \cite{kinconstraint}. Here we have already shown, however, that the new kinetic and derivative interactions are ghost-free by themselves (since individually they {\it are} \EH{} terms up to field re-definitions) and can be part of fully ghost-free spin-2 field theories via the Kinetic/Potential frame versions of spin-2 theories discussed in section \ref{sec-new}.

Several topics suggest themselves for future work. Investigating the background cosmology as well as the behaviour of perturbations in theories involving the new interaction terms will help establish whether new viable cosmological solutions are included within the framework of general spin-2 field theories. In fact, even for exploring already known theories, the Einstein/Jordan and Kinetic/Potential frames described in section \ref{sec-new} should hopefully prove useful. The most pressing question, however, remains what the scale of the ghost in generic spin-2 theories with the new matter couplings and/or the new kinetic and derivative interactions presented here as well as for the non-linear candidate kinetic interactions of \cite{deRham:2013tfa} is -- only once we know this scale will we know the range of validity of general \eft{}s for spin-2 field theories and in which such theories interesting non-linear behaviour can be described. We discuss the scale of all the associated ghosts in \cite{kinconstraint}. In addressing these and other related questions we will hopefully come closer to getting a more complete picture of what consistent theories for spin-2 fields we can build, approaching the same level of understanding we have already obtained for field theories of spin-1 and spin-0 fields.
\\

\noindent {\bf Acknowledgements: } I would like to thank Claudia de Rham, Pedro Ferreira, Andrew Matas, Sigurd N\ae ss, Rachel Rosen, James Scargill, Adam Solomon, David Stefanyszyn and Andrew Tolley for very useful discussions. I am supported by the STFC, BIPAC and the Royal Commission for the Exhibition of 1851. The {\it xAct} package for Mathematica \cite{xAct} was used in the computation and check of some of the results presented here.

\bibliographystyle{JHEP}
\bibliography{kin-bib}

\end{document}